\documentclass[twocolumn]{aastex63}

\usepackage{amssymb,amsmath,amsthm}
\usepackage{graphicx}
\usepackage{mathrsfs}
\usepackage{booktabs}
\usepackage{multirow}
\usepackage{empheq}
\usepackage{verbatim}
\usepackage{mathtools}
\usepackage{hyperref}
\usepackage{xcolor}
\usepackage{xspace}
\usepackage{lineno}

\def\gsim{\mathrel{\raise.3ex\hbox{$>$\kern-.75em\lower1ex\hbox{$\sim$}}}}

\def\deriv{\ensuremath{\text{d}}\xspace} 
\def\uas{\ensuremath{\mu}as\xspace} 
\def\msun{M\ensuremath{_{\odot}}\xspace} 
\def\UM{\textsc{UniverseMachine}\xspace} 

\defcitealias{Narayan_1995b}{NY95}
\defcitealias{Mahadevan_1997}{M97}

\defcitealias{Paper1}{EHTC I}
\defcitealias{Paper2}{EHTC II}
\defcitealias{Paper3}{EHTC III}
\defcitealias{Paper4}{EHTC IV}
\defcitealias{Paper5}{EHTC V}
\defcitealias{Paper6}{EHTC VI}

\begin{document}

\nocite{Paper1}
\nocite{Paper2}
\nocite{Paper3}
\nocite{Paper4}
\nocite{Paper5}
\nocite{Paper6}

\title{Toward determining the number of observable supermassive black hole shadows}

\correspondingauthor{Dominic~W.~Pesce}
\email{dpesce@cfa.harvard.edu}

\author[0000-0002-5278-9221]{Dominic~W.~Pesce}
\affiliation{Center for Astrophysics $|$ Harvard \& Smithsonian, 60 Garden Street, Cambridge, MA 02138, USA}
\affiliation{Black Hole Initiative at Harvard University, 20 Garden Street, Cambridge, MA 02138, USA}

\author[0000-0002-7179-3816]{Daniel~C.~M.~Palumbo}
\affiliation{Center for Astrophysics $|$ Harvard \& Smithsonian, 60 Garden Street, Cambridge, MA 02138, USA}
\affiliation{Black Hole Initiative at Harvard University, 20 Garden Street, Cambridge, MA 02138, USA}

\author[0000-0002-1919-2730]{Ramesh~Narayan}
\affiliation{Center for Astrophysics $|$ Harvard \& Smithsonian, 60 Garden Street, Cambridge, MA 02138, USA}
\affiliation{Black Hole Initiative at Harvard University, 20 Garden Street, Cambridge, MA 02138, USA}

\author[0000-0002-9030-642X]{Lindy~Blackburn}
\affiliation{Center for Astrophysics $|$ Harvard \& Smithsonian, 60 Garden Street, Cambridge, MA 02138, USA}
\affiliation{Black Hole Initiative at Harvard University, 20 Garden Street, Cambridge, MA 02138, USA}

\author[0000-0002-9031-0904]{Sheperd~S.~Doeleman}
\affiliation{Center for Astrophysics $|$ Harvard \& Smithsonian, 60 Garden Street, Cambridge, MA 02138, USA}
\affiliation{Black Hole Initiative at Harvard University, 20 Garden Street, Cambridge, MA 02138, USA}

\author[0000-0002-4120-3029]{Michael~D.~Johnson}
\affiliation{Center for Astrophysics $|$ Harvard \& Smithsonian, 60 Garden Street, Cambridge, MA 02138, USA}
\affiliation{Black Hole Initiative at Harvard University, 20 Garden Street, Cambridge, MA 02138, USA}

\author[0000-0002-4430-102X]{Chung-Pei~Ma}
\affiliation{Department of Astronomy, University of California, Berkeley, CA 94720, USA}
\affiliation{Department of Physics, University of California, Berkeley, CA 94720, USA}

\author[0000-0001-6920-662X]{Neil~M.~Nagar}
\affiliation{Astronomy Department, Universidad de Concepci\'on, Casilla 160-C, Concepci\'on, Chile}

\author[0000-0002-5554-8896]{Priyamvada~Natarajan}
\affiliation{Department of Astronomy, Yale University, 52 Hillhouse Avenue, New Haven, CT 06511, USA}
\affiliation{Department of Physics, Yale University, P.O. Box 208121, New Haven, CT 06520, USA}
\affiliation{Black Hole Initiative at Harvard University, 20 Garden Street, Cambridge, MA 02138, USA}

\author[0000-0001-5287-0452]{Angelo~Ricarte}
\affiliation{Center for Astrophysics $|$ Harvard \& Smithsonian, 60 Garden Street, Cambridge, MA 02138, USA}
\affiliation{Black Hole Initiative at Harvard University, 20 Garden Street, Cambridge, MA 02138, USA}

\begin{abstract}
We present estimates for the number of shadow-resolved supermassive black hole (SMBH) systems that can be detected using radio interferometers, as a function of angular resolution, flux density sensitivity, and observing frequency.  Accounting for the distribution of SMBHs across mass, redshift, and accretion rate, we use a new semi-analytic spectral energy distribution model to derive the number of SMBHs with detectable and optically thin horizon-scale emission.  We demonstrate that (sub)millimeter interferometric observations with ${\sim}0.1$\,\uas resolution and ${\sim}1\,\mu{\rm Jy}$ sensitivity could access ${>}10^6$ SMBH shadows.  We then further decompose the shadow source counts into the number of black holes for which we could expect to observe the first- and second-order lensed photon rings. Accessing the bulk population of first-order photon rings requires ${\lesssim}2$\,\uas resolution and ${\lesssim}0.5$\,mJy sensitivity, while doing the same for second-order photon rings requires ${\lesssim}0.1$\,\uas resolution and ${\lesssim}5$\,$\mu$Jy sensitivity. Our model predicts that with modest improvements to sensitivity, as many as $\sim$5 additional horizon-resolved sources should become accessible to the current Event Horizon Telescope (EHT), while a next-generation EHT observing at 345\,GHz should have access to ${\sim}$3 times as many sources.  More generally, our results can help guide enhancements of current arrays and specifications for future interferometric experiments that aim to spatially resolve a large population of SMBH shadows or higher-order photon rings.
\end{abstract}


\section{Introduction}

The observations and resulting images of the supermassive black hole (SMBH) in the M87 galaxy by the Event Horizon Telescope (EHT) collaboration \citep{Paper1,Paper2,Paper3,Paper4,Paper5,Paper6} represent the first steps in a new field of spatially resolved horizon-scale studies of black holes.  The emission from around the SMBH in M87 takes the form of a bright ring surrounding a darker central ``shadow,'' as expected from simple models of spherical accretion \citep{Falcke_2000,Narayan_2019}.  A wide variety of simulated images of black hole accretion flows show that this ring generically has a diameter that is comparable to the theoretical curve bounding the photon capture cross-section of the time-reversed black hole \citep{Paper5,Paper6}. General relativity predicts that the boundary of this cross-section should take on a nearly circular shape with a diameter of approximately five times the Schwarzschild radius \citep{Bardeen_1973}, and that this diameter should depend only weakly (to within $\pm {\sim}$4\%) on the black hole's spin and inclination \citep{Takahashi_2004,Johannsen_2010}.  These properties permit spatially resolved observations to constrain the black hole mass using measurements of the shadow size; EHT observations of M87 yielded a ${\sim} 10\%$ mass measurement via this approach \citep{Paper6}.

Though the EHT has focused its attention thus far on only those black holes with the largest angular sizes as seen from Earth, almost all massive galaxies are expected to host SMBHs \citep{Magorrian_1998,Kormendy_2013}.  As the EHT and future facilities improve upon the angular resolution and flux density sensitivity of the first M87 observations, more SMBH shadows -- and their corresponding constraints on the black hole masses -- will become observationally accessible.  Though new black hole mass measurements are valuable for individual galaxy studies, questions about SMBH formation and growth mechanisms and the degree to which they co-evolve with their host galaxies are most effectively addressed using large statistical samples of precisely-measured SMBH masses \citep{Volonteri_2010,Heckman_2014}.  To this end, it is natural to ask what observational requirements would be necessary to access large numbers of SMBHs with spatially resolved shadows.

In addition to mass measurements, sufficiently high-resolution observations of SMBHs can also provide unique access to the black hole spin and potentially other spacetime properties.  Hidden within the ring of emission seen by the EHT is an unresolved series of approximately concentric ``photon rings,'' formed by rays that execute increasingly many orbits about the black hole prior to escaping \citep{Darwin_1959,Luminet_1979,Gralla_2019,Johnson_2020}.  Each higher order photon ring -- enumerated by the number $n$ of half-orbits that the constituent photon trajectories make around the black hole -- is expected to have an exponentially narrower angular width on the sky than the previous order.  The lowest-order ($n=0$, corresponding to direct emission) photon ring is impacted by specific details of the accretion flow (e.g., stochastic turbulent structure) that complicate precise spacetime constraints, while the geometric properties of higher-order rings contain the same spacetime information while being exponentially less impacted by such ``astrophysical'' contamination.  Furthermore, interferometric observations naturally decompose the emission by spatial scale, meaning that with fine enough angular resolution the signal from $n > 0$ will dominate the interferometric response in a time-averaged image \citep{Johnson_2020,Gelles_2021}.

The goal of this paper is to determine the number of SMBH shadows and low-order photon rings that could be observed as a function of angular resolution, flux density sensitivity, and observing frequency.  We assume that such observations will be carried out using (sub)millimeter-wavelength interferometry, and we take 230\,GHz to be a characteristic observing frequency when not otherwise specified.  In \autoref{sec:PopSourceCounts} we describe our formalism and input assumptions, which we use to compute the number and distribution of SMBH shadows in the universe as seen from Earth.  In \autoref{sec:SourceCountsInt} we modify these shadow counts to reflect the flux density response expected when observing with interferometers, and we further decompose the total source counts into contributions from systems for which we could observe the $n \geq 0$, $n \geq 1$, and $n \geq 2$ photon rings.  In \autoref{sec:Discussion} we discuss the implications of the source count distributions for current and future telescope specifications.  We summarize and conclude in \autoref{sec:Summary}.  Throughout this paper we assume a flat cosmology with $\Omega_m = 0.3$, $\Omega_{\Lambda} = 0.7$, and $H_0 = 70$\,km\,s$^{-1}$\,Mpc$^{-1}$ unless otherwise specified.

\section{Population source counts} \label{sec:PopSourceCounts}

Our goal is to estimate the number of black hole shadows that we could hope to observe.  Concretely, we would like to determine the number $N(\theta_r,\sigma_{\nu})$ of SMBHs that satisfy the following three conditions:

\begin{enumerate}
    \item The shadow of the black hole has an angular size larger than some resolution threshold $\theta_r$.
    \item The flux density of the horizon-scale emission exceeds some sensitivity threshold $\sigma_{\nu}$.
    \item The emitting plasma is optically thin.
\end{enumerate}

\noindent The first of these criteria is set primarily by the mass of and distance to the black hole, while the second two also depend on the mass accretion rate and the physical conditions in the accretion flow.  The third criterion exists to ensure that we could identify a black hole shadow as such; i.e., an optically thick emission region could obscure the shadow even if the angular resolution and sensitivity would otherwise make it accessible.

\subsection{Overview of strategy} \label{sec:StrategyOverview}

Our strategy for determining $N(\theta_r,\sigma_{\nu})$ starts by considering the global distribution of SMBHs as a function of mass $M$ and redshift $z$,

\begin{equation}
\Phi(M,z) \equiv \frac{dN}{dz\,dM} \label{eqn:BHMFDef}
\end{equation}

\noindent to which we then sequentially apply the above three criteria to narrow down the number of potentially detectable sources.  The distribution $\Phi(M,z)$ is described by the black hole mass function (BHMF), which we discuss in \autoref{sec:BHMF}.

For a given SMBH mass $M$ and redshift $z$, applying our first criterion -- i.e., that the angular shadow size $\vartheta$ is larger than some resolution $\theta_r$ -- amounts to requiring that the black hole mass exceed some minimum mass $m_0(z)$.  A black hole of mass $M$ situated at an angular diameter distance $ D_{\text{A}}$ has an angular shadow size that is given by 

\begin{equation}
\vartheta \approx \sqrt{27} \frac{R_{\text{S}}}{D_{\text{A}}} = \frac{2 \sqrt{27} G M}{c^2 D_{\text{A}}} , \label{eqn:BHDiameter}
\end{equation}

\noindent where $R_{\text{S}}$ is the Schwarzschild radius and the numerical prefactor $\sqrt{27}$ is determined by the shadow diameter for a Schwarzschild black hole \citep{Hilbert1917a,Bardeen_1973}. At a particular redshift $z$, the condition $\vartheta \geq \theta_{r}$ corresponds to

\begin{equation}
M \geq m_0(\theta_r) \equiv \frac{\theta_{r} c^2 D(z)}{2 \sqrt{27} G \left( 1 + z \right)} , \label{eqn:MassThreshold}
\end{equation}

\noindent where $m_0(\theta_r)$ is the critical mass for which a SMBH at redshift $z$ has a shadow with angular size $\theta_r$, and where we have cast the expression in terms of the comoving distance, $D(z) = (1+z) D_{\text{A}}$ \citep[see also][]{Bisnovatyi-Kogan_2018}.

Applying our second condition -- i.e., that the flux density $S_{\nu}$ be greater than some threshold $\sigma_{\nu}$ -- requires knowing the distribution $P(S_{\nu}|M,z)$ of flux densities for a SMBH of mass $M$ at redshift $z$.  The flux density $S_{\nu}(\nu_0)$ observed at a frequency $\nu_0$ is related to the emitted luminosity density $L_{\nu}$ by \citep{Peacock_1999},

\begin{equation}
S_{\nu}(\nu_0) = \frac{L_{\nu}([1+z] \nu_0)}{4 \pi (1 + z) D(z)^2} . \label{eqn:FluxDensity}
\end{equation}

\noindent Here, $L_{\nu}([1+z] \nu_0)$ denotes the luminosity density evaluated at the redshifted frequency $(1+z) \nu_0$, and we have assumed that the emission is isotropic.\footnote{This isotropy assumption is justified because the total flux in the lensed horizon-scale emission from a SMBH accretion flow is not expected to have a strong directional dependence in the same manner as Doppler-boosted jet emission would.}  $L_{\nu}$ is determined by the spectral energy distribution (SED) of the source, which we model as described in \autoref{sec:SED} (with more comprehensive details provided in \autoref{app:SEDmodel}).  Within our SED model, $L_{\nu}$ depends not only on the mass of the SMBH but also on its mass accretion rate $\dot{M}$, which we cast in terms of the Eddington ratio $\lambda$,

\begin{equation}
\lambda \equiv \frac{\dot{M}}{\dot{M}_{\text{Edd}}} .
\end{equation}

\noindent Here, $\dot{M}_{\text{Edd}} \equiv L_{\text{Edd}} / \eta c^2$ is the Eddington mass accretion rate, and $\eta$ is a nominal radiative efficiency that relates $\dot{M}_{\text{Edd}}$ to the Eddington luminosity $L_{\text{Edd}}$; for this paper, we take the radiative efficiency to be $\eta = 0.1$ \citep[e.g.,][]{Yuan_2014}.  Determining $P(S_{\nu}|M,z)$ thus further requires knowledge of the Eddington ratio distribution function (ERDF), which we describe in \autoref{sec:ERDF}.

Applying our third condition -- i.e., that the horizon-scale emission be optically thin at the observing frequency $\nu_0$ -- can also be achieved using our SED model, which provides an optical depth prediction for a SMBH with any given $M$, $\lambda$, and $\nu_0$.  Practically, we can absorb this condition into the definition of the flux density distribution by considering only those systems that are optically thin, i.e., by determining $P(S_{\nu}|M,z,\tau \leq 1)$.  The fraction $f(\sigma_{\nu})$ of SMBHs for which we could expect to detect the horizon-scale emission is then given by

\begin{equation}
f(\sigma_{\nu}) = \int_{\sigma_{\nu}}^{\infty} P(S_{\nu} | M, z, \tau \leq 1) \,\deriv S_{\nu} , \label{eqn:FracSensThresh}
\end{equation}

\noindent where $\sigma_{\nu}$ is some specified sensitivity threshold.

Combining all three criteria, we can compute the source counts expected for any choice of $\theta_r$ and $\sigma_{\nu}$ by integrating the global distribution over mass and redshift,

\begin{equation}
N(\theta_{r},\sigma_{\nu}) = \int_0^{\infty} \deriv z \int_{m_0(\theta_r)}^{\infty} f(\sigma_{\nu}) \Phi(M,z) \,\deriv M . \label{eqn:SourceCountInitial}
\end{equation}

\begin{figure*}
    \centering
    \includegraphics[width=1.00\textwidth]{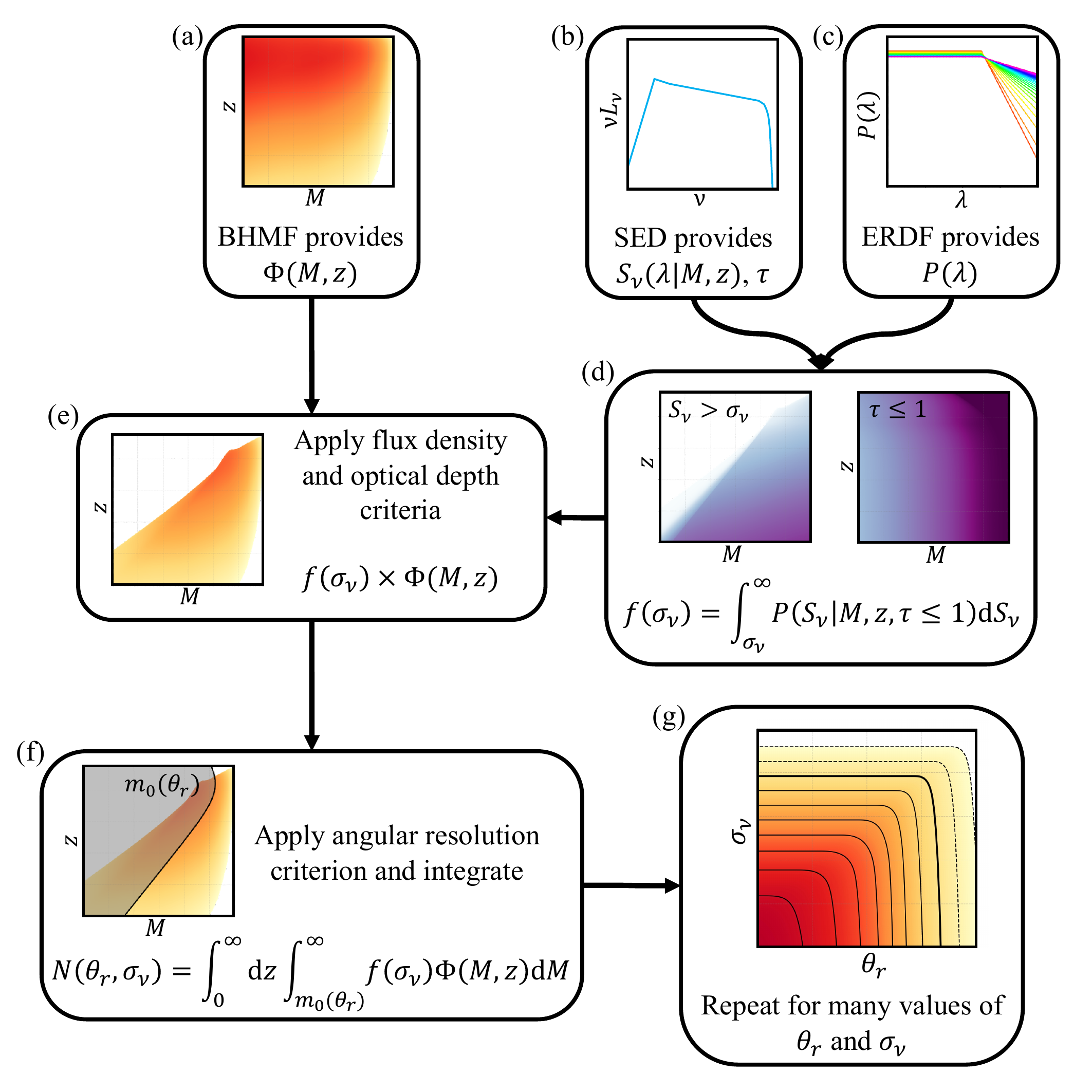}
    \caption{Flowchart illustrating the strategy for determining $N(\theta_r,\sigma_{\nu})$ (see \autoref{sec:StrategyOverview}) using an example case of $\theta_r = 1$\,\uas and $\sigma_{\nu} = 10^{-5}$\,Jy. Panels (a), (b), and (c) show the three primary inputs: the black hole mass function, the SED model, and the Eddington ratio distribution function, respectively.  The BHMF provides the global distribution of SMBHs as a function of $M$ and $z$ (see \autoref{sec:BHMF}), the SED model predicts the emitted flux density and optical depth for every $M$, $\lambda$, and $z$ (see \autoref{sec:SED}), and the ERDF provides the distribution of Eddington ratios $\lambda$ (see \autoref{sec:ERDF}).  In panel (d) the ERDF and the SED model are used to determine the fraction $f(\sigma_{\nu})$ of sources that simultaneously have flux densities exceeding $\sigma_{\nu}$ (left plot in the panel) and are optically thin (i.e., $\tau < 1$; right plot in the panel); in both plots, darker colors indicate a larger fraction.  The combined fraction, as a function of $M$ and $z$, is then used in panel (e) to modify the global SMBH distribution from the BHMF.  In panel (f) we further apply the requirement that the angular shadow size exceed $\theta_r$, which can be cast as a minimum mass $m_0(\theta_r)$ at every $z$; $N(\theta_r,\sigma_{\nu})$ is then determined by integrating over the region outside of the gray shaded area.  Finally, panel (g) illustrates that this procedure can be repeated for many other values of both $\theta_r$ and $\sigma_{\nu}$ (see \autoref{sec:NumberOfShadows}).}
    \label{fig:sequence}
\end{figure*}

\noindent Many of the results presented in this paper are derived from evaluating  \autoref{eqn:SourceCountInitial}.  When computing this integral, we must keep in mind that $m_0(\theta_r)$ is a function of $z$ and that $f(\sigma_{\nu})$ is a function of both $z$ and $M$.  \autoref{fig:sequence} illustrates the procedure we follow to determine $N(\theta_r,\sigma_{\nu})$ for an example set of angular resolution and flux density thresholds (in this case, $\theta_r = 1$\,\uas and $\sigma_{\nu} = 10^{-5}$\,Jy).

\subsection{Black hole mass function} \label{sec:BHMF}

Any evaluation of \autoref{eqn:SourceCountInitial} requires a choice of BHMF, which commonly takes the form

\begin{equation}
\Phi' = \frac{dN}{dV dM} ,
\end{equation}

\noindent where $dN$ is the number of SMBHs in the mass range $(M,M+dM)$ and the comoving volume range $(V,V+dV)$.\footnote{We note that some authors define the BHMF per unit logarithmic (base-10) mass bin, such that their distribution $\phi'$ is related to the one we use by $\phi' = \ln(10) M \Phi'$.}  For our purposes it is more useful to work with $\Phi(M,z)$, the number of black holes in the redshift range $(z,z+dz)$ (see \autoref{eqn:BHMFDef}), which is related to $\Phi'$ by

\begin{eqnarray}
\Phi(M,z) & = & \Phi' \frac{dV}{dz} \nonumber \\
& = & \frac{4 \pi c D^2 \Phi'}{H_0 E(z)} .
\end{eqnarray}

\noindent Here, $E(z) = H(z)/H_0 = \sqrt{\Omega_m (1 + z)^3 + \Omega_{\Lambda}}$ is the dimensionless Hubble parameter \citep{Peebles_1993}.

Estimating the BHMF from observations is difficult because astronomical surveys are inevitably incomplete in ways that impose poorly-known selection functions on the SMBH count in any mass bin, and because there are currently no SMBH mass measurement techniques that are both precise and broadly-applicable \citep{Kelly_2012}.  Many variants of the BHMF thus exist in the literature \citep[e.g.,][]{Salucci_1999,Aller_2002,Marconi_2004,Greene_2007,Lauer_2007,Natarajan_2009,Kelly_2013}.  Recognizing that no single one of these BHMFs is likely to be uniquely correct, in this paper we consider two different BHMF prescriptions -- which we will refer to as our ``lower'' and ``upper'' BHMFs -- that aim to capture a reasonable range of possibilities.

\begin{figure*}[t]
    \centering
    \includegraphics[width=1.00\textwidth]{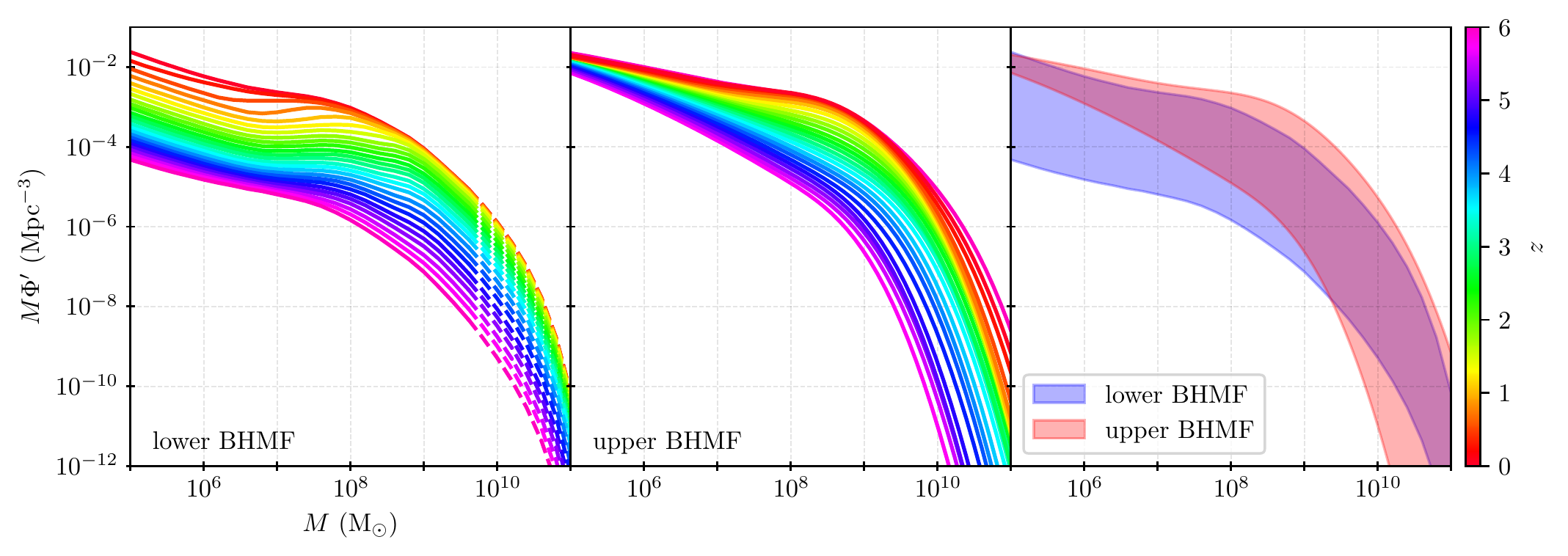}
    \caption{Black hole mass functions used in this paper; see \autoref{sec:BHMF} for details about each BHMF.  The left panel shows the lower BHMF from \cite{Shankar_2009}, with our extrapolation to higher SMBH masses (i.e., $M > 10^{9.5}$\,\msun, as proposed by \citealt{Natarajan_2009}) shown using dashed lines; the BHMF is colored by redshift.  The center panel shows the upper BHMF constructed using the \UM SMF with the \cite{Kormendy_2013} scaling relation.  The right panel shows the envelopes (between $z=0$ and $z=6$) of the BHMFs from the left and middle panels, overlaid to ease comparison.  All three panels share the same horizontal and vertical axis ranges, which are explicitly labeled in the left panel.}
    \label{fig:BHMFs}
\end{figure*}

We take as our lower BHMF the phenomenological model developed by \cite{Shankar_2009} and shown in the left panel of \autoref{fig:BHMFs}.  This BHMF is evolved self-consistently forward in time within a continuity equation formalism \citep{Cavaliere_1971,Small_1992} tuned to match an estimate of the bolometric AGN luminosity function based primarily on the X-ray observations compiled by \cite{Ueda_2003}.  The \cite{Shankar_2009} BHMF is a function of both $M$ and $z$, covering SMBH masses in the range $10^5$--$10^{9.5}$\,\msun and redshifts in the range 0--6.  To account for the known existence of SMBHs with masses exceeding $10^{9.5}$\,\msun \citep[e.g.,][]{Paper6}, we extrapolate the BHMF using a power law with an exponential cutoff,

\begin{equation}
\Phi^{\prime}_{\text{extrapolated}} \propto M^{-a} \exp\left( - \frac{M}{M_{\text{cutoff}}} \right) .
\end{equation}

\noindent The index and normalization of the power law are determined for every $z$ by fitting the BHMF values between $10^9$\,\msun and $10^{9.5}$\,\msun.  \citet{Natarajan_2009} argued on empirical and theoretical grounds for the existence of an upper mass limit for SMBHs at every cosmic epoch.  First, using a physical argument based on self-regulation, they showed that when the accretion energy of a growing SMBH back-reacts with the gas flow and exceeds the binding energy of the feeding disk, it leads to the BH stunting its own growth and results in an upper limit for its mass \citep[see also][]{King_2016}. Empirically, such a limit is expected from the observed SMBH mass - bulge luminosity relation when the relation is extrapolated to the bulge luminosities of bright central galaxies in clusters \citep{Magorrian_1998}.  \citet{Natarajan_2009} showed that consistency between the optical and X-ray BHMFs requires an upper mass limit for local SMBHs that is on the order of ${\sim}10^{10}$\,\msun.  Calibrating their estimates using the more recent observational measurements of the ${\sim}1.7 \times 10^{10}$\,\msun SMBH in NGC 1600 \citep{Thomas_2016}, we determine an exponential cutoff mass of $3.5 \times 10^{10}$\,\msun.  The extrapolated portion of the BHMF is plotted using dashed lines in the left panel of \autoref{fig:BHMFs}.

As a counterpart to the model-based lower BHMF, we also consider an upper BHMF derived empirically using the \UM stellar mass function (SMF) from \cite{Behroozi_2019}.  The \UM SMF is constructed as part of a comprehensive model for galaxy growth spanning redshifts $0 \leq z \leq 10$ and accommodating many observational constraints, including among them a number of observational SMFs determined in various bands \citep{Baldry_2012,Ilbert_2013,Moustakas_2013,Muzzin_2013,Tomczak_2014,Song_2016}. From the \UM SMF, we convert from stellar mass $M_*$ to SMBH mass $M$ using the scaling law from \cite{Kormendy_2013} as done in \cite{Ricarte_2018},

\begin{equation}
\log\left( \frac{M}{\text{ M}_{\odot}} \right) = 8.69 + 1.16 \log\left( \frac{M_*}{10^{11} \text{ M}_{\odot}} \right) . \label{eqn:KHConversion}
\end{equation}

\noindent After converting from stellar to SMBH mass, we convolve the SMBH mass distributions with a Gaussian kernel with a 0.3-dex FWHM to account for the intrinsic scatter in the scaling relations.  The resulting upper BHMF is shown in the center panel of \autoref{fig:BHMFs}.

Relative to the lower BHMF, the upper BHMF predicts systematically more SMBHs at low to intermediate redshifts (i.e., $z \lesssim 3$) and at all masses, though at the highest redshifts the lower BHMF predicts more SMBHs with $M \gtrsim 10^{9.5}$\,\msun (see right panel of \autoref{fig:BHMFs}).  The low-redshift behavior of the lower BHMF agrees well with a BHMF derived from the \UM~SMF using the \cite{McConnell_2013} scaling relation \citep[see also][]{Saglia_2016}\footnote{An even lower BHMF could be produced using, e.g., the scaling relation from \cite{Reines_2015}, but the resulting BHMF systematically underpredicts the observed local Universe's high-mass SMBH population by several orders of magnitude.}.  To remain conservative in our estimates, throughout this paper we treat the lower BHMF as our fiducial case and use it for all computations and figures unless otherwise specified; we use the upper BHMF primarily to determine plausible uncertainty ranges for computed values.  For this paper, we treat both BHMFs as being nonzero only in the range $0 \leq z \leq 6$ and $10^5 \leq M \leq 10^{11}$\,\msun.

For the analyses carried out in this paper, the high-mass end of the BHMF is most important.  To assess the fidelity of the high-mass end of the lower and upper BHMFs, we compare their predictions against the number of known massive SMBHs in the local universe.  In this regard, the MASSIVE galaxy survey provides a convenient comparison point because it is a volume-limited survey targeting massive early-type galaxies with stellar masses above $10^{11.5}$\,\msun within a distance of 108\,Mpc, or $z \approx 0.02$ \citep{Ma_2014}.  To date, four\footnote{This number should be taken as a lower limit because the MASSIVE survey is ongoing and may uncover more SMBHs in the same range of $M$ and $z$.} SMBHs in this volume have dynamically measured masses at or above M87’s $M \geq 6.5 \times 10^9$\,\msun:  M87 \citep{Paper6}, NGC 1600 \citep{Thomas_2016}, NGC 3842, and NGC 4889 \citep{McConnell_2011}.  Our lower and upper BHMFs predict that the number of SMBHs within $z \leq 0.02$ and $M > 6.5 \times 10^9$\,\msun should be $\sim$5 and $\sim$29, respectively, which are consistent with the MASSIVE survey results.  The specific behavior of the BHMF at low masses is less important because these black holes do not contribute significantly at the angular resolutions and flux densities of most interest for this paper.

\subsection{Spectral energy distribution model} \label{sec:SED}

Given the global distribution of SMBHs across mass and redshift, \autoref{eqn:SourceCountInitial} selects only the fraction $f(\sigma_{\nu})$ that have optically thin emission with a flux density that exceeds the sensitivity threshold.  This fraction is defined in \autoref{eqn:FracSensThresh}, and it results from integrating over the distribution of flux densities $P(S_{\nu}|M,z)$ at a given $M$ and $z$.  The first piece of information we need to compute this integral is an SED model, which will permit us to determine the flux density $S_{\nu}(\lambda | M,z)$ corresponding to a particular choice of Eddington ratio, black hole mass, and redshift, and also to assess when the observed emission will be optically thin.

Observational constraints on SMBH growth indicate that SMBHs spend the majority of their time accreting at well below the Eddington rate \citep{Hopkins_2006c}.  At these low accretion rates, the material in the vicinity of the black hole is thought to follow the advection-dominated accretion flow (ADAF) solution to the hydrodynamic equations describing viscous and differentially-rotating flows around black holes \citep{Narayan_1995b,Narayan_1998,Yuan_2014}.  An ADAF accretion disk has a two-temperature structure in which the ion temperature is greater than the electron temperature.  The electrons are able to cool via a combination of synchrotron, bremsstrahlung, and inverse Compton radiation, which together define the SED for the observed emission.

For SMBHs observed in the radio to submillimeter wavelength range, as relevant for this work, the SED is dominated by synchrotron and Compton emission.  \citet[][hereafter \citetalias{Mahadevan_1997}]{Mahadevan_1997} provides a convenient formalism for computing the gross spectral properties of an ADAF system given a black hole mass $M$ and accretion rate $\dot{M}$ \citep[see also][]{Narayan_1994,Narayan_1995a,Narayan_1995b}.  We use a modified version of the \citetalias{Mahadevan_1997} formalism for the SED models in this paper, and \autoref{app:SEDmodel} provides a detailed description of our updated model.  We note that this SED model only considers emission from the accretion flow, and it does not incorporate a jet component.

Our SED model provides an estimate of the emitted luminosity density $L_{\nu}$ as a function of frequency for any input values of $M$ and $\lambda$.  Given a particular redshift $z$, we convert $L_{\nu}$ to $S_{\nu}$ using \autoref{eqn:FluxDensity}.  We determine whether the system is optically thin by comparing the rest-frame observing frequency, $(1+z) \nu_0$, to the peak synchrotron frequency in the source, $\nu_p$ (see \autoref{eqn:CriticalSynchFreq}).  So long as $(1+z) \nu_0 \geq \nu_p$, we consider the system to be optically thin.

\subsection{Eddington ratio distribution function} \label{sec:ERDF}

The last piece of information we need to compute the integral in \autoref{eqn:FracSensThresh} is an ERDF, which provides a probabilistic description of what fraction of SMBHs should be accreting at any particular Eddington rate $\lambda$.  In this paper, we consider every SMBH to be active at some level, rather than considering the accretion to have only binary ``on'' and ``off'' states.  We thus dispense with the notion of a ``duty cycle'' often adopted for AGN (or equivalently, we take the duty cycle to be unity), and we instead work exclusively in terms of an Eddington ratio distribution function \citep[e.g.,][]{Merloni_2008} to account for the differences in accretion rates.

\begin{figure}
    \centering
    \includegraphics[width=1.00\columnwidth]{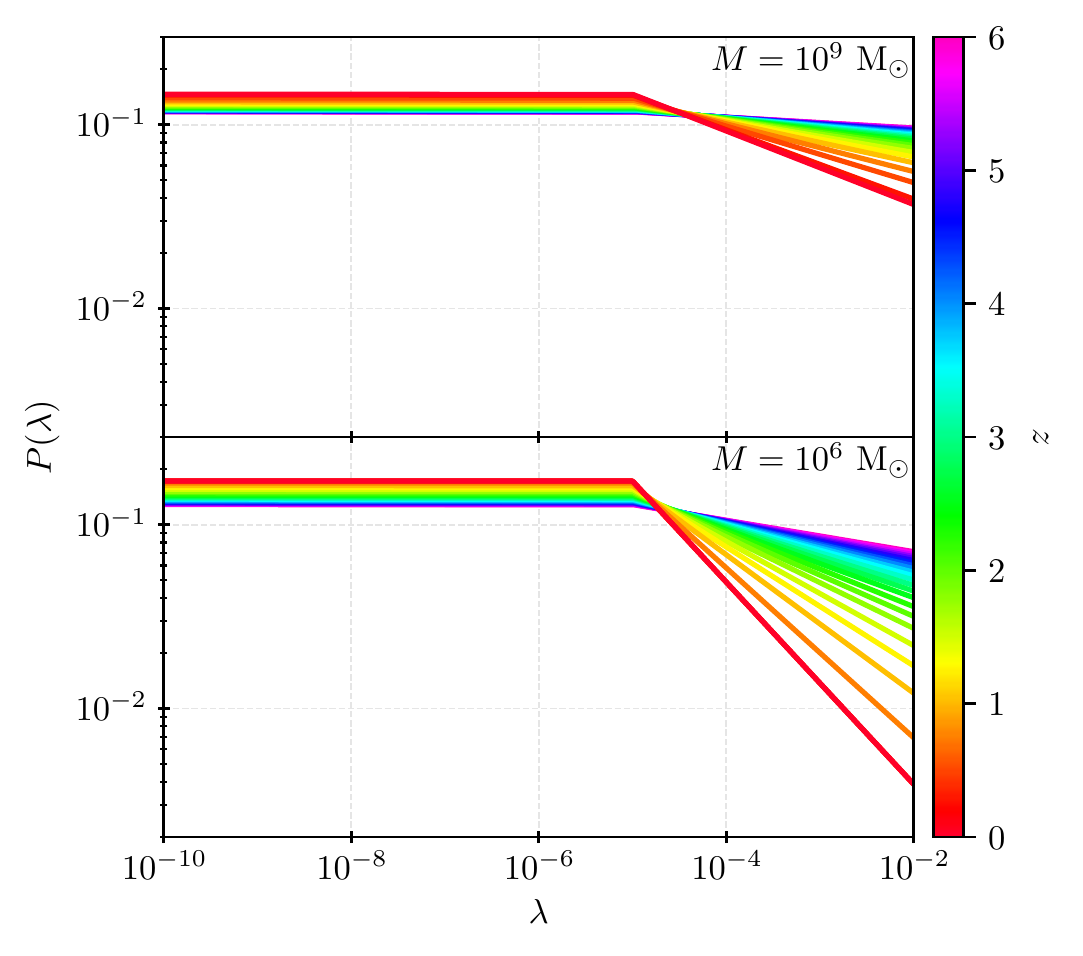}
    \caption{Eddington ratio distribution function, adapted from \cite{Tucci_2017} and updated using the measurements from \cite{Aird_2018}; see \autoref{sec:ERDF} for details. The top panel shows the ERDF plotted as the probability $P(\lambda)$ per unit $\log(\lambda)$ as a function of $\lambda$ and $z$ for a SMBH mass of $M = 10^9$\,\msun, and the bottom panel shows the same for a SMBH mass of $M = 10^6$\,\msun.}
    \label{fig:ERDF}
\end{figure}

There is emerging evidence that luminous (``Type 1''; unobscured; $\lambda \gtrsim 10^{-2}$) and low-luminosity (``Type 2''; obscured; $\lambda \lesssim 10^{-2}$) AGN follow different distributions \citep{Kauffmann_2009,Trump_2011,Weigel_2017}.  Though the ERDF for luminous AGN appears to be consistent with a log-normal distribution \citep{Lusso_2012}, there is no clear consensus in the literature on a specific form for the ERDF of low-lumminosity AGN (LLAGN).  Different authors have used variants that include a power-law \citep{Aird_2012,Bongiorno_2012}, a Schechter function \citep{Hopkins_2009,Cao_2010,Hickox_2014}, and a log-normal \citep{Kauffmann_2009,Conroy_2013}.  Additionally, while there seems to be broad agreement on a power-law behavior towards low Eddingtion ratios in the local Universe (i.e., $z \lesssim 1$), few observational constraints currently exist for the ERDF of LLAGN at $z \gtrsim 1$.

We proceed with a form for the ERDF adapted from the analytic prescription used by \cite{Tucci_2017} and updated using the more recent measurements from \cite{Aird_2018}.  For their ERDF, \cite{Tucci_2017} used a Schechter function with an exponential cutoff value of $\lambda = 1.5$, but for our purposes (i.e., LLAGN with $\lambda \ll 1)$ only the power-law component of the ERDF is relevant.  Furthermore, the LLAGN portion of the ERDF from \cite{Tucci_2017} was constructed to match the low-redshift behavior from \cite{Hopkins_2009}, \cite{Kauffmann_2009}, and \cite{Aird_2012}.  None of these previous papers included observational constraints for AGN accreting below $\lambda \approx 10^{-5}$.  To avoid the strong dependence on the low-end cutoff that comes from continuing the power law to arbitrarily small values, we posit instead that the distribution breaks \citep[as in, e.g.,][]{Weigel_2017}.  Specifically, we modify the power-law ERDF from \cite{Tucci_2017} such that it flattens out for Eddington ratios smaller than some value $\lambda_0$.  That is, we have

\begin{equation}
P(\lambda) = \begin{cases}
A , & \lambda_{\text{min}} \leq \lambda \leq \lambda_0 \\
A \left( \frac{\lambda}{\lambda_0} \right)^{\alpha} , & \lambda_0 < \lambda < \lambda_{\text{max}}
\end{cases} , \label{eqn:ProbDistEdd}
\end{equation}

\noindent where $P(\lambda)$ is the probability density per unit logarithmic interval in $\lambda$, $\lambda_{\text{min}}$ and $\lambda_{\text{max}}$ are the lowest and highest permitted values, and the coefficient $A$ is constructed such that the distribution integrates to unity:

\begin{equation}
A = \left[ \log\left( \frac{\lambda_0}{\lambda_{\text{min}}} \right) + \frac{1}{\alpha \lambda_0^{\alpha} \ln(10)} \left( \lambda_{\text{max}}^{\alpha} - \lambda_0^{\alpha} \right) \right]^{-1} .
\end{equation}

\noindent In this paper, we use values of $\lambda_{\text{min}} = 10^{-10}$, $\lambda_0 = 10^{-5}$, and $\lambda_{\text{max}} = 10^{-2}$ (see \autoref{sec:CriticalMdot}).

In addition to permitting the power-law index $\alpha$ to evolve with redshift, we also allow for additional evolution with SMBH mass,

\begin{equation}
\alpha(z,M) = a(M) \times \begin{cases}
- 1 , & z \leq a(M) \\
- 1 / \left[ 1 + z - a(M) \right] , & z > a(M)
\end{cases} . \label{eqn:ERDFalpha}
\end{equation}

\noindent Here, $a(M)$ encodes the mass dependence of the power-law index.  Though there is some prior observational evidence indicating that the ERDF is approximately independent of SMBH mass \citep{Kauffmann_2009,Kelly_2013,Weigel_2017}, recent measurements by \cite{Aird_2018} found that more massive SMBHs tend to be accreting at higher rates.  We thus treat $a(M)$ as being essentially bimodal, with low-mass SMBHs having one power-law index value and high-mass SMBHs having another, and we use a logistic function to smoothly vary $a(M)$ between these two extremes,

\begin{equation}
a(M) = \frac{a_{\text{hi}} + a_{\text{lo}} \left( M / M_0 \right)^{- 1 / \Delta}}{1 + \left( M / M_0 \right)^{- 1 / \Delta}} . \label{eqn:ERDFa}
\end{equation}

\noindent Here, $a_{\text{lo}}$ describes the power-law index at small masses, $a_{\text{hi}}$ describes the power-law index at large masses, $M_0$ denotes the midpoint mass, and $\Delta$ is the logistic width in $\log(M)$ that controls how quickly the transition from the low-mass regime to the high-mass regime occurs.  We determine the values of these four parameters by fitting \autoref{eqn:ProbDistEdd} to the \cite{Aird_2018} measurements; our fitting procedure is described in \autoref{app:ERDFMassDependence}.  We find best-fit values of $a_{\text{lo}} = 0.55$, $a_{\text{hi}} = 0.20$, $\log(M_0) = 7.5$, and $\Delta = 0.3$, and the resulting ERDF is shown in \autoref{fig:ERDF}.

\autoref{eqn:ProbDistEdd} defines the probability $P(\lambda)$ per unit $\log(\lambda)$ for any particular SMBH to be accreting at the rate $\lambda$.  Given some specified $M$ and $z$, we determine the probability $P(S_{\nu}|M,z)$ by numerically sampling from $P(\lambda)$ and using our SED model (see \autoref{sec:SED}) to associate each sample with a particular $S_{\nu}$.  Efficient sampling of $P(\lambda)$ can be achieved by transforming a random variable $x$ that is distributed according to a unit uniform distribution through the inverse cumulative distribution function (CDF) of \autoref{eqn:ProbDistEdd}.  This inverse CDF is given by

\footnotesize
\begin{align}
& \text{CDF}^{-1}(x) = \nonumber \\
& \begin{cases}
\lambda_{\text{min}} 10^{x/A} , & 0 \leq x \leq A \log\left( \frac{\lambda_0}{\lambda_{\text{min}}} \right) \\
\lambda_0 \left( \alpha \ln(10) \left[ \frac{x}{A} - \log\left( \frac{\lambda_0}{\lambda_{\text{min}}} \right) \right] + 1 \right)^{1/\alpha} , & A \log\left( \frac{\lambda_0}{\lambda_{\text{min}}} \right) < x \leq 1
\end{cases} ,
\end{align}
\normalsize

\noindent which we can use to generate random samples distributed according to \autoref{eqn:ProbDistEdd}.  The associated distribution of $S_{\nu}$ provides an estimate of $P(S_{\nu} | M,z)$, which we then integrate per \autoref{eqn:FracSensThresh} for the purposes of evaluating \autoref{eqn:SourceCountInitial}.

\subsection{The number of black hole shadows} \label{sec:NumberOfShadows}

\begin{figure}
    \centering
    \includegraphics[width=1.00\columnwidth]{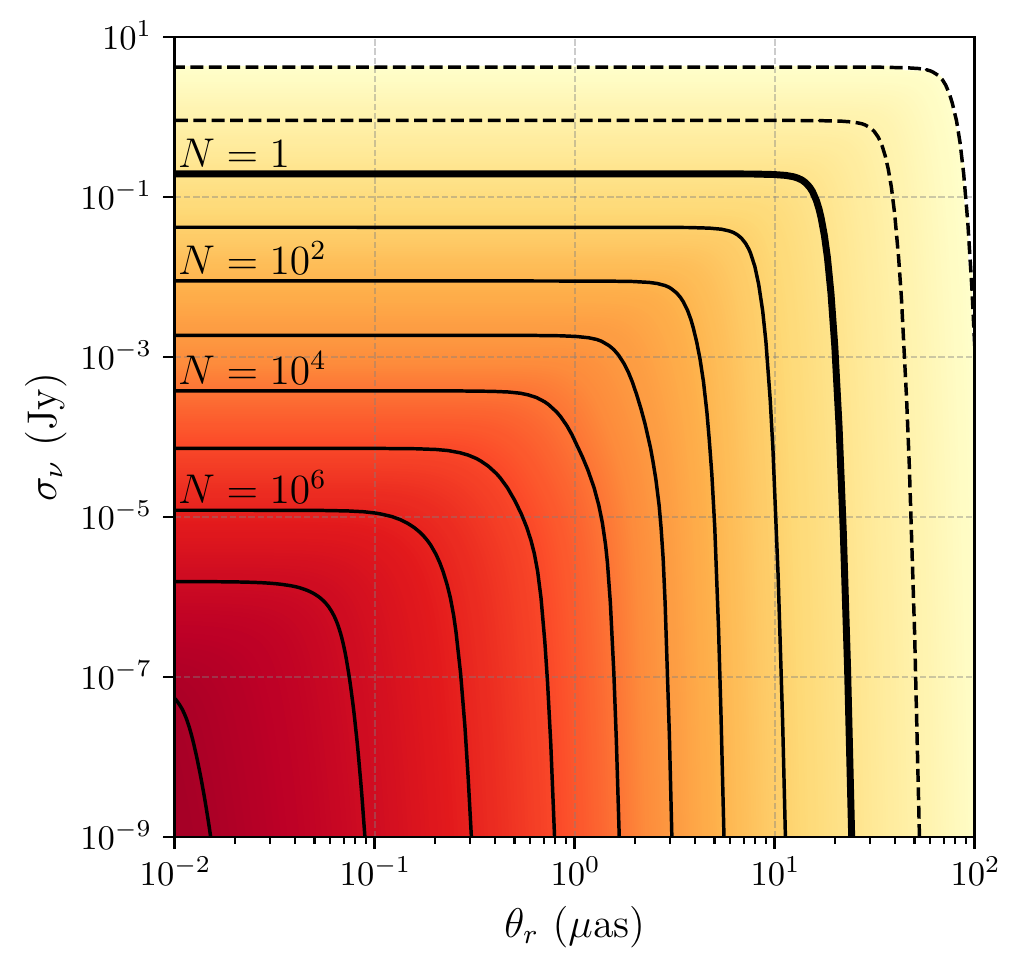}
    \includegraphics[width=1.00\columnwidth]{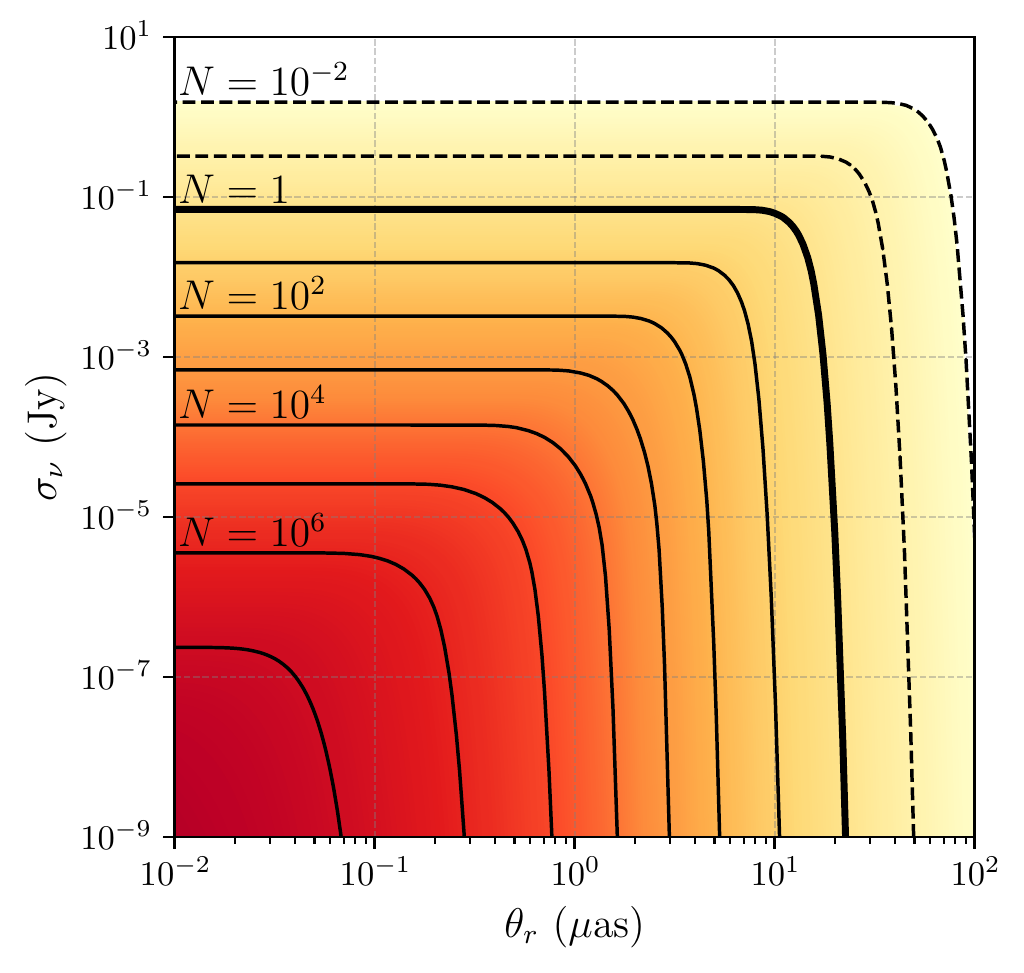}
    \caption{\textit{Top}: Number of black hole shadows with angular sizes larger than an angular resolution threshold $\theta_r$ and total flux densities larger than a sensitivity threshold $\sigma_{\nu}$, as a function of the threshold values and assuming an observing frequency of 230 GHz; i.e., $N(\theta_r,\sigma_{\nu})$ from \autoref{eqn:SourceCountInitial}.  The solid contours start with the thick contour indicating a count of $N=1$ and then increase by factors of 10 towards the lower left, while the dashed contours each decrease by a factor of ten towards the upper right. \textit{Bottom}: Same as the top panel, but with the additional restriction that the sources must be optically thin.}
    \label{fig:source_counts_230GHz_taucut}
\end{figure}

Putting it all together, \autoref{fig:source_counts_230GHz_taucut} shows the result of evaluating \autoref{eqn:SourceCountInitial} over a range of values for both the angular resolution threshold $\theta_r$ and the flux density sensitivity $\sigma_{\nu}$ at an observing frequency of $\nu_0 = 230$\,GHz.  The top panel shows the source counts predicted without imposing the optical depth condition, while the bottom panel restricts the sources to those that satisfy $\tau \leq 1$ (see \autoref{eqn:FracSensThresh}).  Each point in both panels of \autoref{fig:source_counts_230GHz_taucut} is computed from an integral over the remaining $(M,z)$ space.  These plots thus represent an observation-independent prediction about the character of the SMBH population; namely, how many SMBHs are expected to have angular shadow sizes in excess of $\theta_r$, horizon-scale flux densities at 230\,GHz greater than $\sigma_{\nu}$, and (in the case of the bottom panel) an optically thin accretion flow.  An approximate analytic description of the resulting $N(\theta_r,\sigma_{\nu})$ is provided in \autoref{app:AnalyticApprox}.

\begin{figure*}
    \centering
    \includegraphics[width=1.00\textwidth]{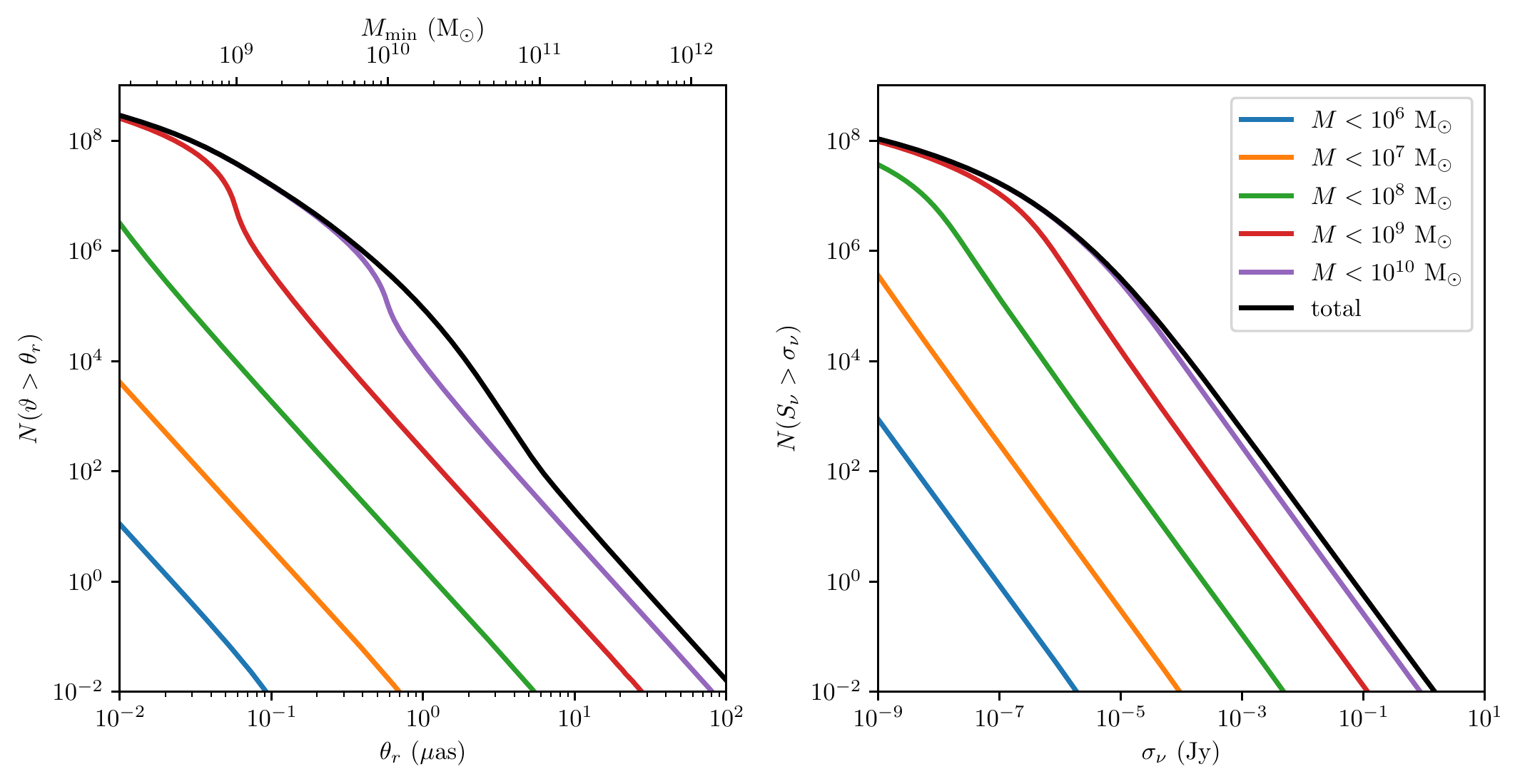}
    \caption{One-dimensional slices through $N(\theta_r,\sigma_{\nu})$ for an observing frequency of 230\,GHz, with no restriction on the optical depth $\tau$. \textit{Left}: The black curve shows the total number of SMBHs with shadows larger than some threshold angular resolution $\theta_r$ as a function of that threshold; this curve approximately corresponds to a horizontal cut through the bottom part of the top panel of \autoref{fig:source_counts_230GHz_taucut}.  The upper axis indicates the minimum mass of a black hole for which the corresponding angular resolution would permit that black hole to be spatially resolved at any redshift. \textit{Right}: The black curve shows the total number of SMBHs with horizon-scale flux densities larger than some threshold value $\sigma_{\nu}$ as a function of that threshold; this curve approximately corresponds to a vertical cut through the left part of the top panel of \autoref{fig:source_counts_230GHz_taucut}.  In both panels the source counts for different choices of black hole mass binning are shown as colored curves.}
    \label{fig:1D_slices}
\end{figure*}

The two panels of \autoref{fig:1D_slices} show the behavior of $N(\theta_r,\sigma_{\nu})$ in the limit as $\sigma_{\nu} = 0$ (left panel) and $\theta_r = 0$ (right panel); these limits correspond approximately to one-dimensional slices through the top panel of \autoref{fig:source_counts_230GHz_taucut} along the horizontal and vertical axes, respectively.  The black curve in the left panel shows $N(\vartheta > \theta_r, \sigma_{\nu}=0)$, while the colored curves show the contribution from SMBHs in different mass ranges.  At large $\theta_r$ we see that the source counts follow the $N \propto \theta_r^{-3}$ behavior expected from simple volume scaling.  The upturn around $\theta_r \approx 1$\,\uas occurs because this is the resolution threshold below which the most massive SMBHs can be seen at any redshift (because of the turnover in angular diameter distance at $z \approx 1.6$), and the re-flattening at smaller $\theta_r $ is caused by the finite redshift coverage of the BHMF.  The black curve in the right panel shows $N(\theta_r = 0, S_{\nu} > \sigma_{\nu})$, while the colored curves again split out the contribution by SMBH mass.  Throughout most of the space we see the source counts climbing volumetrically as the flux density decreases, following $N \propto \sigma_{\nu}^{-3/2}$.  Cosmological effects become noticeable at the lowest $\sigma_{\nu}$ values, where the curve starts flattening out owing to a combination of the luminosity distance increasing more rapidly as well as the finite redshift coverage of the BHMF.

\section{Interferometric source counts} \label{sec:SourceCountsInt}

The analysis performed in the previous section predicts the source counts corresponding to the population of SMBHs that adhere to the three criteria specified at the beginning of \autoref{sec:PopSourceCounts}.  We now aim to estimate a subtly different quantity: the number of shadow-resolved sources that could be observed by a telescope with angular resolution $\theta_r$ and flux density sensitivity $\sigma_{\nu}$.  This conceptual distinction is relevant because the telescopes that we expect to be carrying out spatially resolved studies of black hole shadows in the foreseeable future are radio interferometers.  While the source counting analysis performed in \autoref{sec:PopSourceCounts} uses the SED model detailed in \autoref{app:SEDmodel} to determine the flux density expected from any particular SMBH, this SED model only provides an estimate for the total (i.e., spatially integrated) horizon-scale flux density.  However, an interferometric baseline is only sensitive to flux on specific spatial scales, determined by the length of the baseline and the wavelength of light being observed.  Though in this paper we do not explore specific methods for estimating shadow diameters, sparse interferometric observations have previously been used to constrain the shadow diameter for M87 under the assumption that the source is ring-like \citep{Doeleman_2012,Wielgus_2020}.  In this section, we thus investigate the prospects for detecting SMBH shadows on an individual interferometric baseline.

\subsection{Flux density seen by a single baseline} \label{sec:SingleBaseline}

We base our expectations for the horizon-scale emission structure from a SMBH on the observational and theoretical understanding of the M87 system. \cite{Johnson_2020} provide an approximate analytic expression for the expected flux density of the photon ring emission as a function of baseline length for optically thin emission, which we adapt to take the following form:

\begin{align}
S(u) = \eta S_0 J_0(\pi \vartheta u) \sum_{n=0}^{\infty} e^{-n \pi} e^{-\frac{\left(\pi u W_n\right)^2}{4\ln(2)}} , \label{eqn:RingVisApprox}
\end{align}

\begin{equation}
W_n \approx W_0 e^{-n \pi} . \label{eqn:SubringWidth}
\end{equation}

\begin{figure}
    \centering
    \includegraphics[width=1.00\columnwidth]{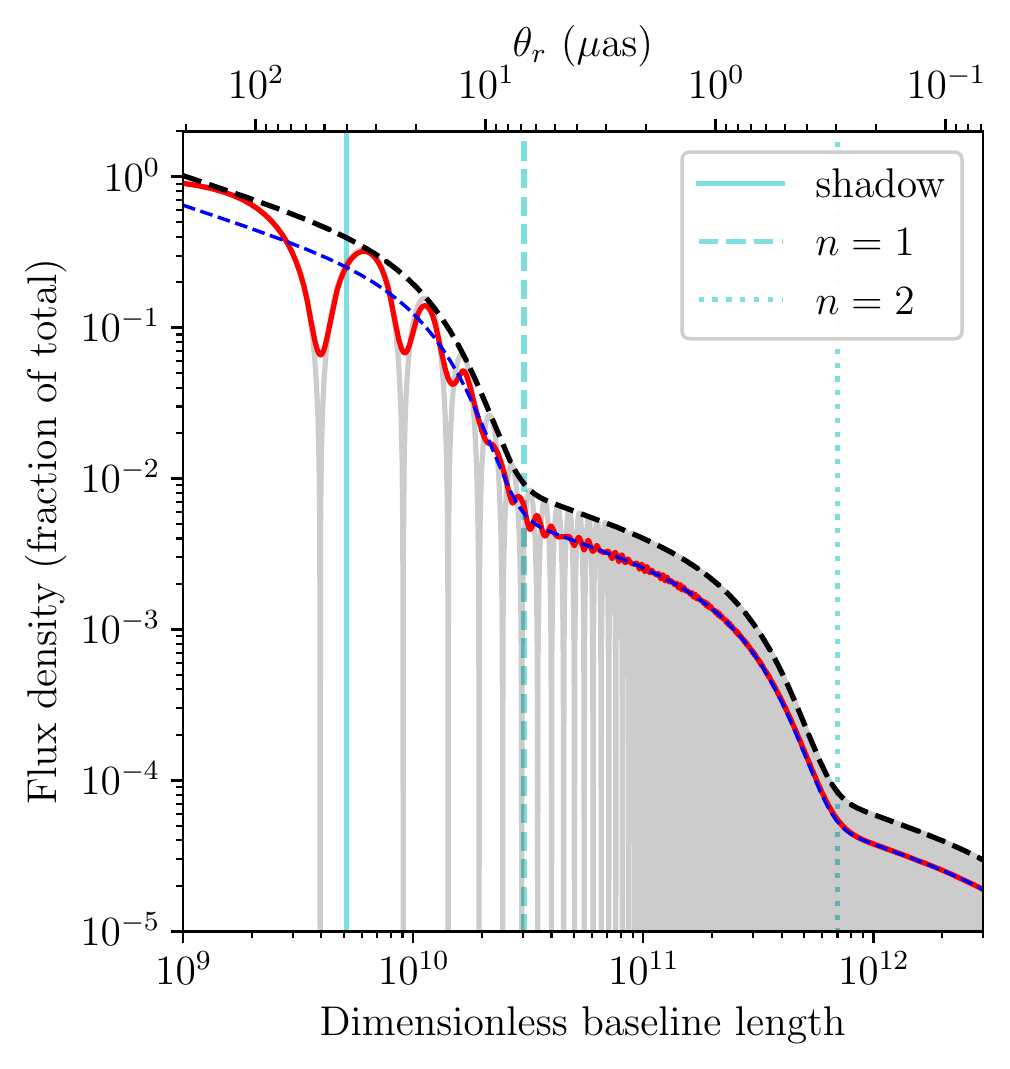}
    \caption{Fraction of the total source flux density that can be detected on long baselines for the photon ring model described in \autoref{sec:SingleBaseline}, shown here for a $\vartheta = 40$\,\uas diameter.  The gray curve shows \autoref{eqn:RingVisApprox}, and the dashed black line tracks the envelope of this function.  The red curve shows a running average of the gray curve across a 2\% fractional observing bandwidth, and the dashed blue curve shows \autoref{eqn:BWAvgFluxDens}. The vertical cyan lines show the resolution criteria used for the shadow ($\theta_r = \vartheta$) and for the first two orders of photon ring ($\theta_r = 2 w_{n-1}$).}
    \label{fig:photon_ring_uv}
\end{figure}

\noindent Here, $S_0$ is the total flux density (i.e., the value provided by the SED model, given the redshift of the SMBH), $\vartheta$ is the angular diameter of the photon ring (which for our purposes is given by \autoref{eqn:BHDiameter}), $u$ is the length of the baseline in units of wavelengths, $W_0$ is the FWHM angular thickness of the lowest-order (i.e., $n=0$) photon ring, and $\eta = 1 - e^{-\pi}$ is a normalizing prefactor.  We assume $W_0 = \vartheta / 5$ \citep{Paper5,Paper6}. \autoref{eqn:RingVisApprox} is shown as the gray curve in \autoref{fig:photon_ring_uv}.

\begin{figure}
    \centering
    \includegraphics[width=1.00\columnwidth]{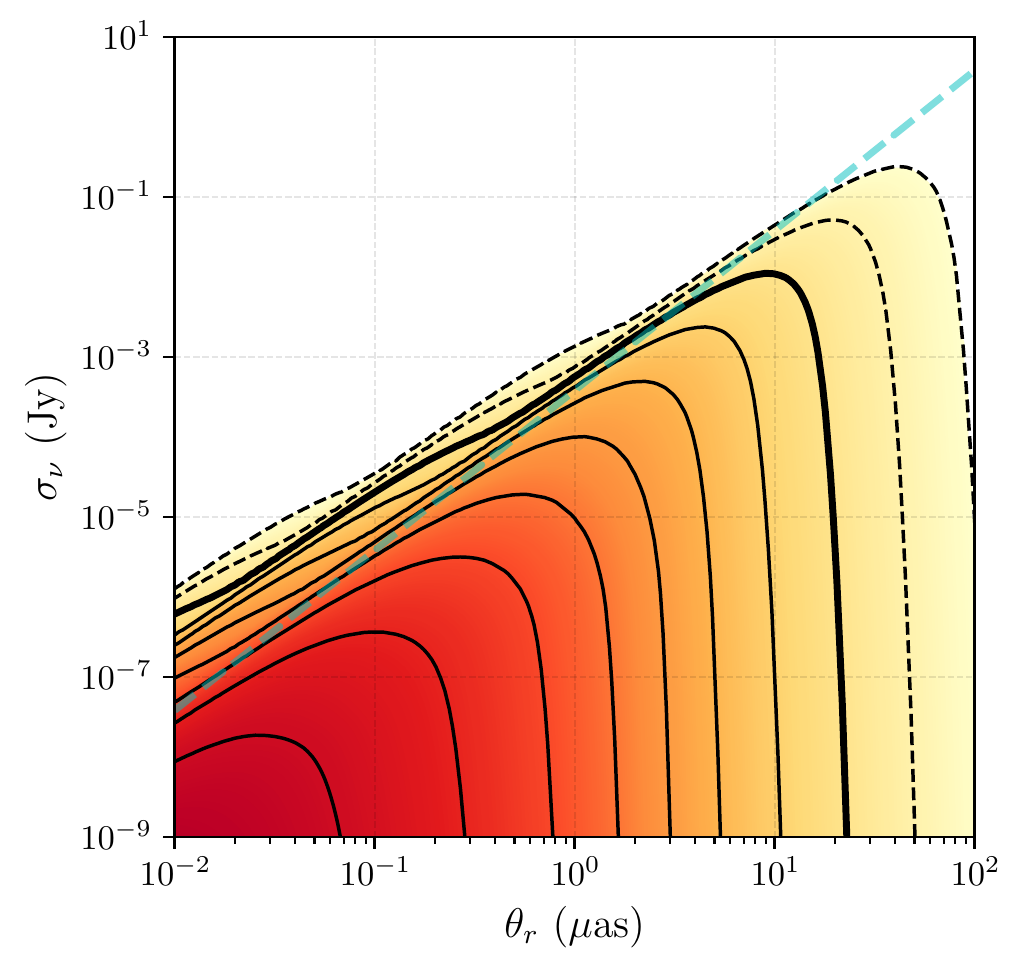}
    \caption{Similar to the bottom panel of \autoref{fig:source_counts_230GHz_taucut}, but now showing the number of shadow-resolved, optically thin SMBHs that could be observed at 230\,GHz by an interferometric baseline with flux density sensitivity $\sigma_{\nu}$ and angular resolution $\theta_r$ across the entire sky.  The drawn contours mark the same source count values as those in \autoref{fig:source_counts_230GHz_taucut}.  The diagonal dashed cyan line marks a constant brightness temperature of $10^{10}$\,K.}
    \label{fig:source_counts_230GHz_taucut_rescut}
\end{figure}

On long baselines (i.e., $u \gg 1/\vartheta$), the bandwidth-averaged flux density will be given by

\begin{equation}
\bar{S}(u) \approx \frac{2 \eta S_0}{\pi^2} \sqrt{\frac{2}{\vartheta u}} \sum_{n=0}^{\infty} e^{-n \pi} e^{-\frac{\left(\pi u W_n\right)^2}{4\ln(2)}} , \label{eqn:BWAvgFluxDens}
\end{equation}

\noindent which is smaller by a factor $2/\pi$ than the envelope of \autoref{eqn:RingVisApprox} as a result of averaging over many periods; \autoref{eqn:BWAvgFluxDens} is shown as the dashed blue curve in \autoref{fig:photon_ring_uv}.  By replacing $S_{\nu}$ in \autoref{eqn:FracSensThresh} with $\bar{S}(1/\theta_r)$ from \autoref{eqn:BWAvgFluxDens} and then recomputing the SMBH source counts via \autoref{eqn:SourceCountInitial}, the form of $N(\theta_r,\sigma_{\nu})$ becomes that shown in \autoref{fig:source_counts_230GHz_taucut_rescut}.  Unlike in \autoref{fig:source_counts_230GHz_taucut}, the source counts no longer monotonically increase as angular resolution improves (i.e., as $\theta_r$ decreases), because \autoref{eqn:BWAvgFluxDens} ensures that longer baselines see lower flux densities from any given SMBH.  An analytic approximation for the resulting $N(\theta_r,\sigma_{\nu})$ is provided in \autoref{app:AnalyticApprox}.

\begin{figure*}
    \centering
    \includegraphics[width=1.00\textwidth]{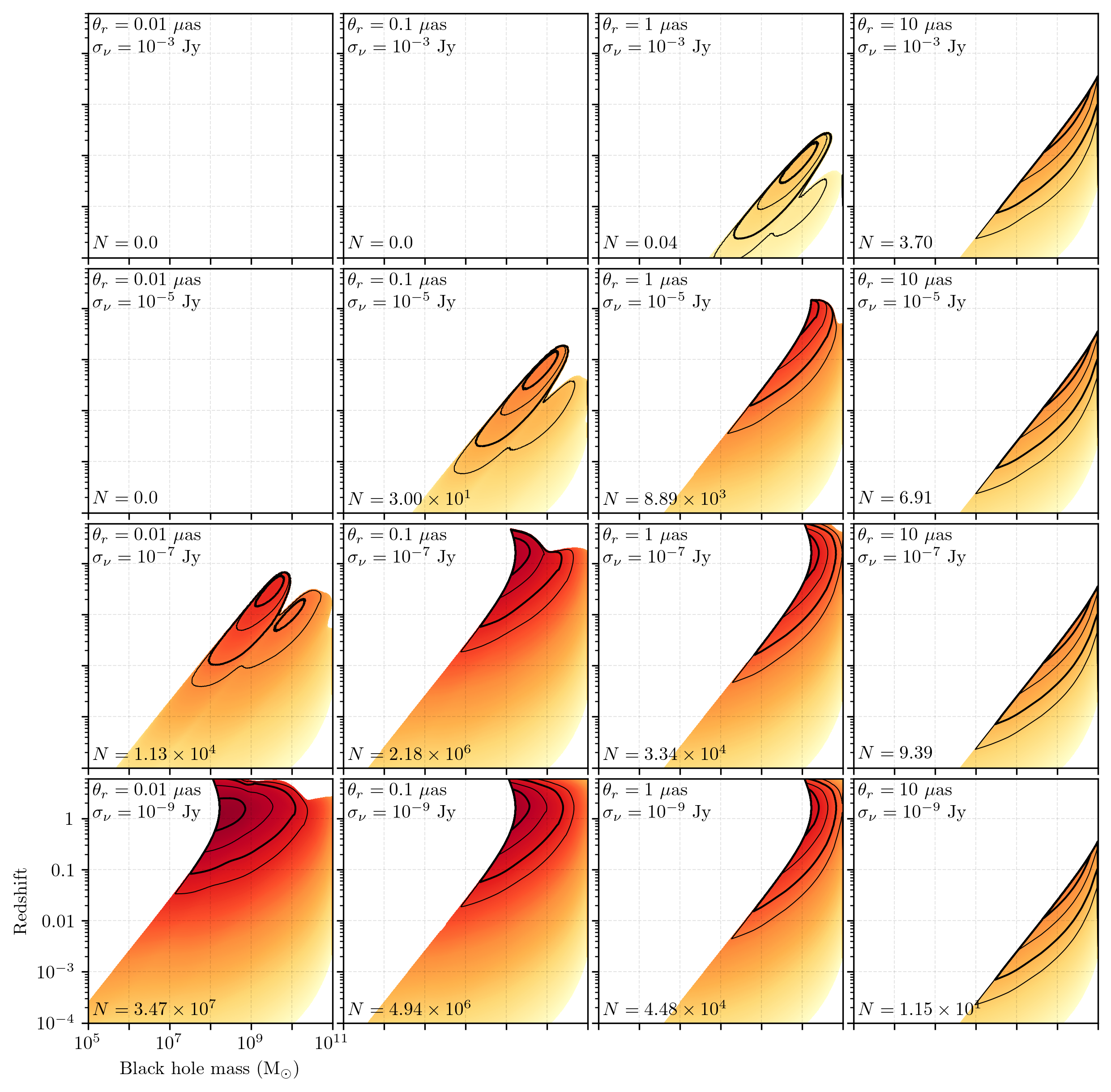}
    \caption{The integrand from \autoref{eqn:SourceCountInitial}, plotted logarithmically as $\frac{dN}{d\ln(M) d\ln(z)}$, showing the distribution of the number of shadow-resolved and optically thin SMBHs that can be seen by a single baseline as a function of redshift and black hole mass. Each panel shows a different choice of $\theta_r$ and $\sigma_{\nu}$, and all panels assume an observing frequency of 230\,GHz.  The total number of black holes, integrated over $M$ and $z$, is given in the lower left-hand corner of each panel.  The colorscale maps to the logarithm of the source number density (i.e., the number of sources per unit logarithmic interval in $M$ and $z$), and the black contours enclose 50\%, 90\%, 99\%, and 99.9\% of the total source count. All panels share the same horizontal and vertical axis ranges, which are explicitly labeled in the bottom-left panel.}
    \label{fig:Mz_distributions_230GHz}
\end{figure*}

\autoref{fig:Mz_distributions_230GHz} shows the integrand from \autoref{eqn:SourceCountInitial} plotted over the domain of integration for several values of $\theta_r$ and $\sigma_{\nu}$, providing the distribution of observable SMBHs as a function of $M$ and $z$.  The number of objects generally increases with increasing redshift (at fixed mass) and with decreasing mass (at fixed redshift), though the density peaks at $z \approx 2$ for the smallest values of $\sigma_{\nu}$.  For certain configurations, such as $\theta_r = 0.1$\,\uas and $\sigma_{\nu} = 10^{-5}$\,Jy, the impact of \autoref{eqn:BWAvgFluxDens} is visually apparent as a lack of monotonicity in the source counts with increasing redshift (at fixed mass).  This behavior reflects the fact that a fixed baseline becomes sensitive to emission from larger spatial scales around a particular SMBH as that SMBH is moved to larger distances; i.e., $\bar{S}$ increases as $\vartheta$ decreases.  On certain intervals in $z$, this flux increase associated with smaller $\vartheta$ is more than sufficient to compensate for the flux decrease associated with the increased distance to the SMBH.

\subsection{Photon ring decomposition} \label{sec:PhotonRingDecomp}

The expression in \autoref{eqn:BWAvgFluxDens} for the horizon-scale flux density contains contributions from all orders of photon rings, and in \autoref{fig:photon_ring_uv} we can see that rings of different order are expected to dominate the observed flux density on different baseline length intervals.  Depending on the value of $u$ relative to $1/\vartheta$, a telescope may thus be primarily sensitive to emission from photon rings with $n > 0$.  To determine the number of sources from which we expect to be able to detect higher-order photon rings, we can decompose the total source counts into bins corresponding to which order of photon ring dominates the emission.

We take as our resolution requirement to ``see'' the $n$th sub-ring that $\theta_r \leq 2 w_{n-1}$, where $w_n = W_n / \sqrt{8 \ln(2)}$ is the Gaussian width corresponding to the FWHM $W_n$ (\autoref{eqn:SubringWidth}).  This angular resolution requirement can be re-cast as a mass threshold $m_n$ for a given redshift, analogous to \autoref{eqn:MassThreshold}; for $n > 0$, we have

\begin{equation}
M \geq m_n \equiv 5 \sqrt{2 \ln(2)} e^{(n-1)\pi} m_0 ,
\end{equation}

\noindent where $m_0$ is defined in \autoref{eqn:MassThreshold}.  \autoref{fig:photon_ring_uv} marks the $n > 0$ and $n > 1$ resolution thresholds using vertical dashed cyan lines.  To ensure that the emission is optically thin enough to see down to the $n$th sub-ring, we further impose a more stringent condition on the optical depth of

\begin{equation}
\tau \leq \tau_n = \frac{1}{n + 1} .
\end{equation}

\noindent By replacing the lower mass limit $m_0$ in \autoref{eqn:SourceCountInitial} with $m_n$, and by replacing the $\tau \leq 1$ condition in \autoref{eqn:FracSensThresh} with $\tau \leq \tau_n$, we can compute the source counts associated with objects for which a photon ring of order $n$ or greater is detectable.

\begin{figure*}
    \centering
    \includegraphics[width=0.9\textwidth]{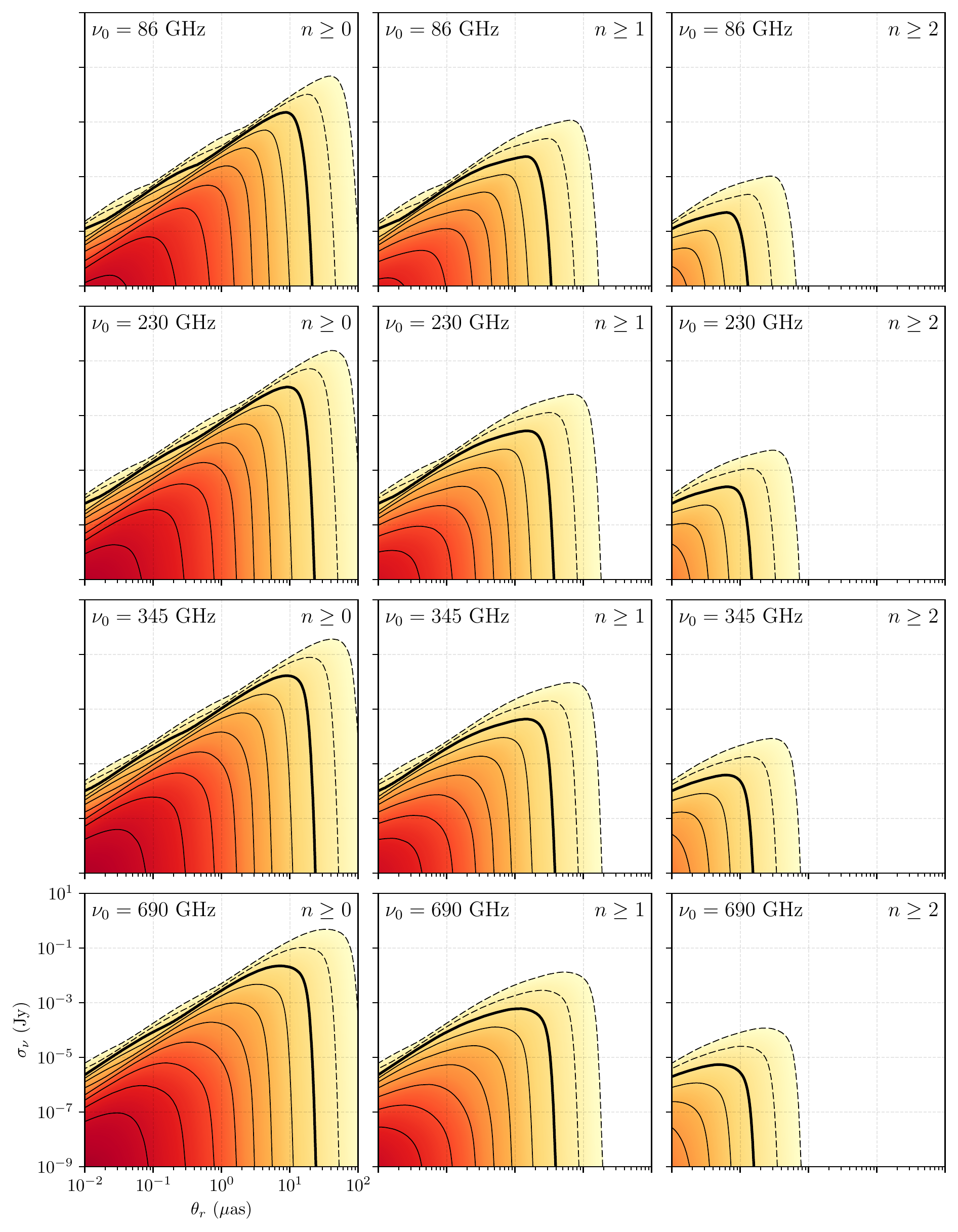}
    \caption{Each panel shows a plot analogous to that in \autoref{fig:source_counts_230GHz_taucut_rescut}, but decomposed into the number of sources for which we could expect to detect the first three orders of photon ring.  Each row shows this decomposition for one of four observing frequencies, with the frequency labeled in the upper left-hand corner of each panel.  For each row of panels, the left panel shows the number of sources for which we could detect any order of photon ring at all, while the center and right panels show the number of sources for which we could detect the first- and second-order photon rings, respectively.  In each panel, the drawn contours mark the same source count values as those in \autoref{fig:source_counts_230GHz_taucut}.  All panels share the same horizontal and vertical axis ranges, which are explicitly labeled in the bottom-left panel.}\label{fig:source_counts_photon_ring_decomp}
\end{figure*}

\autoref{fig:source_counts_photon_ring_decomp} shows these source counts for the first three orders of photon ring at observing frequencies of 86\,GHz, 230\,GHz, 345\,GHz, and 690\,GHz, corresponding to standard atmospheric transmission windows \citep{TMS}.  At each observing frequency we see qualitatively similar behavior: the source counts corresponding to the higher-order photon rings look approximately like scaled-down versions of the $n \geq 0$ counts.  For each additional order, the same source count value is achieved at an angular resolution threshold that is approximately $\sim$20 times finer and a sensitivity threshold that is approximately $\sim$100 times fainter than was necessary at the previous order.  The angular resolution increment is associated with the factor $e^{-\pi} \approx 1/23$ in \autoref{eqn:SubringWidth} that sets the angular size ratio between consecutive photon rings.  The flux density increment comes from a combination of a similar exponential suppression factor (a factor $e^{-\pi}$ from the summand of \autoref{eqn:BWAvgFluxDens}) as well as the fact that the flux density profile is being observed on baselines that are typically a factor of $e^{\pi}$ longer, thereby incurring an additional flux density factor of $e^{-\pi/2} \approx 1/5$ from the $u^{-1/2}$ proportionality in \autoref{eqn:BWAvgFluxDens}.

The evolution of the source counts with frequency primarily affects the required sensitivity, with higher-frequency observations achieving the same source counts at a higher value of $\sigma_{\nu}$ than lower-frequency observations.  The flux density threshold required to detect a particular number of objects is about one order of magnitude smaller at 86\,GHz than at 690\,GHz; i.e., about an order of magnitude better sensitivity -- in terms of Jy -- is required at 86\,GHz than at 690\,GHz.  The angular resolution requirement does not show substantial evolution with frequency across this range.

\subsection{The impact of baseline projection} \label{sec:BaselineProjection}

The analysis presented in this section thus far has assumed that an interferometric baseline can observe the entire sky with the same angular resolution.  However, in reality any physical baseline between two stations will have a different projected length as seen from different locations in the sky.  The resolving power of the baseline will thus be a function of source location on the sky, which means that the number of black hole shadows a baseline can detect per unit solid angle will also vary across the sky.

For a particular baseline, we can define a spherical coordinate system $(\theta,\phi)$ such that $\theta$ is a polar angle measured from the axis defined by the baseline orientation and $\phi$ is measured azimuthally around this axis.  $\theta_r(\theta)$ is then the effective angular resolution of the baseline when projected toward a source at a sky position with polar angle $\theta$,

\begin{equation}
\theta_r (\theta , b) = \frac{1}{b \sin{\theta}} = \frac{\theta_{r,0}}{\sin{\theta}} , \label{eqn:BaselineRes}
\end{equation}

\noindent where $b$ is the baseline length in units of the observing wavelength and $\theta_{r,0} = 1/b$ is the angular resolution achieved when $\theta = \pi/2$ (i.e., the finest resolution achievable by the baseline).  Denoting the number density of sources per unit solid angle as $\frac{d^2N}{d\Omega}$, we can express the total number of sources observable by this baseline as

\begin{equation}
N(b) = \iint_{\Omega_{\text{vis}}} \sin\theta \frac{d^2N}{d\Omega}\left[ \theta_r (\theta , b) \right] \, \deriv \theta \, \deriv \phi , \label{eqn:InstantaneousBaselineSampling}
\end{equation}

\noindent where we have explicitly indicated that the number density is a function of the angular resolution, $\theta_r(\theta , b)$, and we have assumed that sources are distributed isotropically on the sky such that there is no $\phi$ dependence.  The integral is carried out over the solid angle $\Omega_{\text{vis}}$ on the sky that is visible to the baseline.

\begin{figure}
    \centering
    \includegraphics[width=1.00\columnwidth]{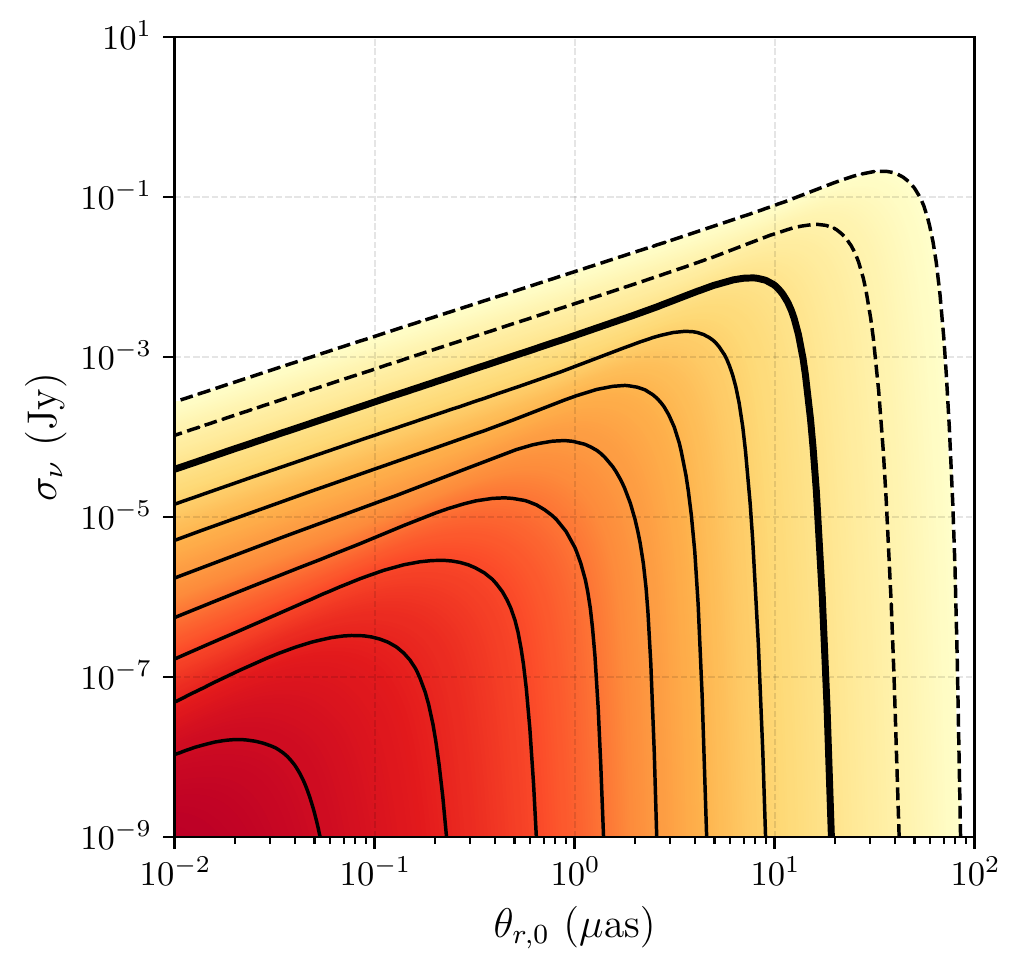}
    \caption{Similar to \autoref{fig:source_counts_230GHz_taucut_rescut}, but now accounting for baseline projection effects appropriate for a space-based interferometric baseline that can see the entire sky (see \autoref{sec:BaselineProjection}).  The colorscale and contours show the number of shadow-resolved, optically thin SMBHs that could be observed at 230\,GHz by an interferometric baseline with finest angular resolution $\theta_{r,0}$ and flux density sensitivity $\sigma_{\nu}$.  The drawn contours mark the same source count values as those in \autoref{fig:source_counts_230GHz_taucut}.}
    \label{fig:source_counts_baseline_projection}
\end{figure}

To illustrate the impact of this geometric effect on source counts, we consider the concrete example of an interferometric baseline formed between two space-based antennas, each of which can see the entire sky.  In this case, the function $d^2N/d\Omega$ is given simply by $N(\theta_r,\sigma_{\nu}) / 4 \pi$, and the domain of integration for \autoref{eqn:InstantaneousBaselineSampling} will be all $(\theta,\phi)$; \autoref{fig:source_counts_baseline_projection} shows the result of this evaluation.  Relative to the source counts in \autoref{fig:source_counts_230GHz_taucut_rescut}, at large values of $\theta_{r,0}$ (e.g., ${\sim}20$\,\uas) the source counts in \autoref{fig:source_counts_baseline_projection} are reduced because some fraction of the sky is not observed with sufficient angular resolution to see SMBHs with shadow sizes that are close to $\theta_{r,0}$.  The magnitude of this reduction is modest, amounting to a factor of $3 \pi / 16 \approx 0.59$ for uniformly distributed sources in flat space (see \autoref{eqn:AnalyticInstantaneousSampling} with $\alpha = 4$).  However, a much more pronounced impact can be seen in the region of fine angular resolution and poor sensitivity (e.g., the region around $\theta_{r,0} \approx 10^{-1}$\,\uas and $\sigma_{\nu} \approx 10^{-4}$\,Jy), where the source counts in \autoref{fig:source_counts_baseline_projection} are significantly increased relative to \autoref{fig:source_counts_230GHz_taucut_rescut}.  This difference arises because $N(\theta_r,\sigma_{\nu})$ climbs rapidly toward larger $\theta_r$ in this region, and so the coarser angular resolutions arising from baseline projection provides access to many SMBHs that a baseline with a fixed angular resolution of $\theta_{r,0}$ across the entire sky would be unable to see.  In this region of the $(\theta_{r,0},\sigma_{\nu})$ space, the impact of baseline projection is to increase the accessible number of SMBH shadows by several orders of magnitude.

While \autoref{eqn:InstantaneousBaselineSampling} provides the source counts appropriate for a fixed baseline, in real-world arrays the baseline will typically be changing orientation with time.  For instance, a spaceborne antenna forming a baseline with another antenna situated on the Earth would execute a complete revolution once every orbital period, as observed by a distant source.  One effect of this rotation is to make a larger fraction of the sky observable with the finest resolution than would otherwise be possible with just the instantaneous configuration, up to a unit fraction if both stations are spaceborne and thus can view the entire sky.  The net impact of rotating the baseline is to bring more SMBH shadows into view than would be accessible by a static baseline.  \autoref{app:RotatingBaseline} provides a more detailed exposition of the sampling behavior of such a baseline as it rotates.

\begin{figure}
    \centering
    \includegraphics[width=1.00\columnwidth]{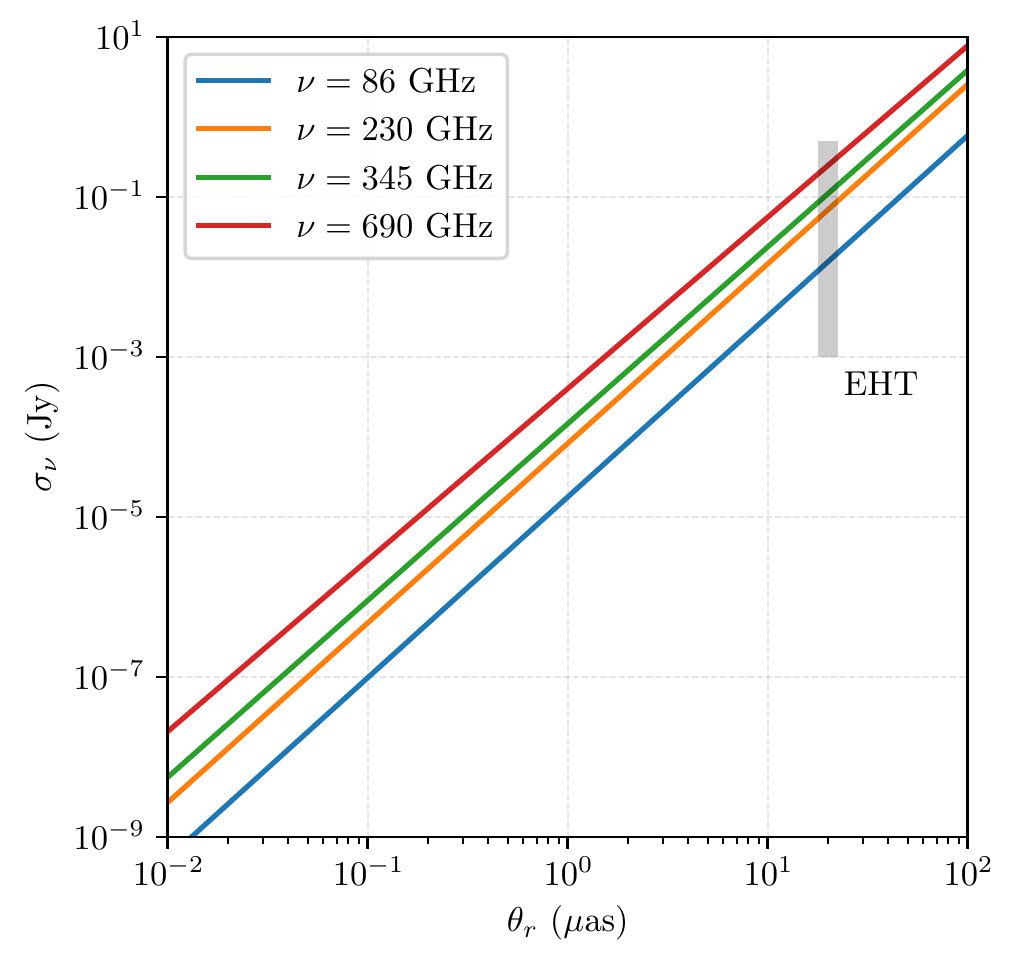}
    \caption{Power-law fits to the ridge-line in $N(\theta_r,\sigma_{\nu})$ -- defined as the location of maximum $\sigma_{\nu}$ for every fixed $N$ -- for four different observing frequencies; this ridge-line can be seen as the turnover in the contours in \autoref{fig:source_counts_230GHz_taucut_rescut}.  Configurations of $(\theta_r,\sigma_{\nu})$ that fall below and to the right of the ridge-line can most effectively increase $N$ by improving angular resolution (i.e., by decreasing $\theta_r$), while configurations that fall above and to the right of the ridge-line can increase $N$ by improving sensitivity (decreasing $\sigma_{\nu}$) or by increasing $\theta_r$.  For reference, we mark the approximate specifications of the EHT (i.e., $\theta_r = 20$\,\uas, $\sigma_{\nu}$ between 1\,mJy and 0.5\,Jy) by a shaded gray region.}\label{fig:pareto_front}
\end{figure}

\begin{deluxetable*}{lccc}
\tablecolumns{4}
\tablewidth{0pt}
\tablecaption{Population characteristics at 230\,GHz\label{tab:PopulationStats}}
\tablehead{  &   & \multicolumn{2}{c}{\textbf{Properties}}\\\cline{3-4}
\textbf{Target population} & \textbf{How many?} & \textbf{$\boldsymbol{\vartheta}$ ($\boldsymbol{\mu}$as)} & \textbf{$\boldsymbol{S_{\nu}}$ (Jy)}}
\startdata
\multirow{5}{*}{black hole shadows} & M87 & 40 & $0.5$ \\
 & $1$ & $(12.8,24.5)$ & $(3.8,16.5) \times 10^{-2}$ \\
 & $10^2$ & $(3.1,5.4)$ & $(1.6,5.5) \times 10^{-3}$ \\
 & $10^4$ & $(0.81,1.1)$ & $(7.9,17.0) \times 10^{-5}$ \\
 & $10^6$ & $(0.15,0.23)$ & $(1.9,4.9) \times 10^{-6}$ \\
\enddata
\tablecomments{Predicted approximate shadow size ($\vartheta$) and 230\,GHz horizon-scale flux density ($S_{\nu}$) above which there exists the listed number of SMBH shadows.  Quantities in parenthesis indicate the values determined from the lower and upper BHMF prescriptions.  We note that our source counting model predicts more stringent requirements to see $N \approx 1$ source than are actually required to see the SMBH in M87 (see \autoref{sec:M87}).  We have thus separately listed the requirements needed to observe M87.}
\end{deluxetable*}

\begin{deluxetable*}{lccc}
\tablecolumns{4}
\tablewidth{0pt}
\tablecaption{Telescope requirements to achieve various source counts at 230\,GHz\label{tab:SourceCounts}}
\tablehead{  &   & \multicolumn{2}{c}{\textbf{Requirements}}\\\cline{3-4}
\textbf{Target population} & \textbf{How many?} & \textbf{$\boldsymbol{\theta_r}$ ($\boldsymbol{\mu}$as)} & \textbf{$\boldsymbol{\sigma_{\nu}}$ (Jy)}}
\startdata
\multirow{5}{*}{black hole shadows ($n \geq 0$)} & M87 & 40 & $10^{-1}$ \\
 & $1$ & $(8.6,16.3)$ & $(1.1,3.9) \times 10^{-2}$ \\
 & $10^2$ & $(2.2,3.5)$ & $(5.0,16.5) \times 10^{-4}$ \\
 & $10^4$ & $(0.5,0.7)$ & $(1.9,5.6) \times 10^{-5}$ \\
 & $10^6$ & $(0.10,0.14)$ & $(3.7,13.0) \times 10^{-7}$ \\
\midrule
\multirow{5}{*}{first-order photon rings  ($n \geq 1$)} & M87 & 7 & $10^{-3}$ \\
 & $1$ & $(1.4,2.6)$ & $(2.7,9.9) \times 10^{-4}$ \\
 & $10^2$ & $(0.31,0.55)$ & $(1.2,4.2) \times 10^{-5}$ \\
 & $10^4$ & $(0.075,0.11)$ & $(5.0,14.6) \times 10^{-7}$ \\
 & $10^6$ & $(0.015,0.022)$ & $(1.2,3.6) \times 10^{-8}$ \\
\midrule
\multirow{3}{*}{second-order photon rings  ($n \geq 2$)} & M87 & 0.3 & $10^{-5}$ \\
 & $1$ & $(0.061,0.12)$ & $(2.5,9.0) \times 10^{-6}$ \\
 & $10^2$ & $(0.013,0.025)$ & $(1.1,3.8) \times 10^{-7}$ \\
\enddata
\tablecomments{Similar to \autoref{tab:PopulationStats}, but listing the predicted approximate single-baseline angular resolution ($\theta_r$) and flux density sensitivity ($\sigma_{\nu}$) requirements for observing different numbers of SMBH shadows and low-order photon rings at 230\,GHz.  Quantities in parenthesis indicate the values determined from the lower and upper BHMF prescriptions.  For each order of photon ring, we have explicitly listed the requirements needed to observe M87 at that order.}
\end{deluxetable*}

\section{Discussion} \label{sec:Discussion}

Our general strategy for carrying out the various source counting analyses presented in this paper is laid out in \autoref{sec:StrategyOverview} and illustrated in \autoref{fig:sequence}.  To recap:

\begin{itemize}
    \item We start with the BHMF, which describes the global distribution $\Phi(M,z)$ of SMBHs across mass and redshift.
    \item Using our SED model and a prescription for the distribution of SMBH accretion rates (i.e., the ERDF), we determine the fraction $f(\sigma_{\nu})$ of objects for which the horizon-scale emission is both optically thin and has either a total flux density (in \autoref{sec:PopSourceCounts}) or a resolved flux density (in \autoref{sec:SourceCountsInt}) exceeding some threshold $\sigma_{\nu}$.
    \item We then integrate the product $f(\sigma_{\nu}) \Phi(M,z)$ over $M$ and $z$, excluding objects with shadow sizes smaller than some angular resolution threshold $\theta_r$ (see \autoref{eqn:SourceCountInitial}).
\end{itemize}

\noindent The quantity $N(\theta_r,\sigma_{\nu})$ resulting from this procedure corresponds to the number of sources with shadow sizes larger than $\theta_r$ and flux densities greater than $\sigma_{\nu}$.

\autoref{fig:source_counts_230GHz_taucut} shows a summary of the SMBH population in terms of the angular shadow size $\vartheta > \theta_r$ and the total horizon-scale flux density $S_{\nu} > \sigma_{\nu}$.  These source counts at any $(\theta_r,\sigma_{\nu})$ provide an estimate for the number of SMBHs that are ``resolvable'' -- i.e., distinguishable from a point source -- by a telescope that achieves an angular resolution of $\theta_r$ and a flux density sensitivity of $\sigma_{\nu}$.  I.e., even if the telescope lacks the sensitivity to detect the source structure on the scale of $\theta_r$, it will still be able to constrain the angular size of the source so long as the telescope's sensitivity is sufficient to detect a total flux density of $\sigma_{\nu}$.\footnote{In practice, an interferometric array carrying out such a measurement will need to have at least a moderately-filled aperture; if instead only a single baseline is present, then the various considerations detailed in \autoref{sec:SourceCountsInt} will apply.}  We find that the population source counts approximately follow the simple scaling relations expected if the number of sources grows with the accessed volume (see \autoref{app:AnalyticApprox}); for example, $\sim$hundreds of sources are predicted to be resolvable with an angular resolution of $\sim$1\,\uas and a flux density sensitivity of $\sim$1\,mJy.

For interferometric observations, we find that the number of detectable SMBH shadows generally increases as the angular resolution $\theta_r$ and sensitivity $\sigma_{\nu}$ improve, but that the gradient of $N(\theta_r,\sigma_{\nu})$ changes orientation throughout the parameter space (see \autoref{fig:source_counts_230GHz_taucut_rescut}).  At large $\theta_r$ and small $\sigma_{\nu}$, the source counts increase exclusively toward smaller $\theta_r$; at large $\sigma_{\nu}$ and small $\theta_r$, the source counts increase both toward smaller $\sigma_{\nu}$ and toward larger $\theta_r$.  The gradient changes orientation from pointing primarily toward smaller $\theta_r$ to pointing primarily toward smaller $\sigma_{\nu}$ around a ridge-line in the $(\theta_r,\sigma_{\nu})$ space that approximately follows a power law $\sigma_{\nu} \propto \theta_r^{2.2}$; \autoref{fig:pareto_front} shows power-law fits to this ridge-line for four different observing frequencies. At an observing frequency of 230\,GHz, we find a best-fit power law of

\begin{equation}
\left( \frac{\sigma_{\nu}}{15 \text{ mJy}} \right) \approx \left( \frac{\theta_r}{10 \text{ \uas}} \right)^{2.2} . \label{eqn:Pareto}
\end{equation}

\noindent This expression can be used to estimate the angular resolution and sensitivity corresponding to an effective ``Pareto front''\footnote{A ``Pareto front'' is the set of locations within a space of interest that satisfy the property that no one condition can be relaxed without making another more stringent.  In our case, the ``Pareto front'' constitutes the set of locations in $(\theta_r,\sigma_{\nu})$ space where neither the angular resolution threshold nor the flux density threshold can be increased (i.e., made less demanding) without requiring a decrease in the other, while still being sensitive to the same number of objects.} in source counts, whereby $(\theta_r,\sigma_{\nu})$ pairs living on this curve are in some sense maximally economical.  That is, to access the same number of sources using a different set of $(\theta_r,\sigma_{\nu})$ would require improving either the sensitivity or the angular resolution.  \autoref{tab:PopulationStats} provides estimates for the number of SBHMs with shadow sizes and optically thin horizon-scale flux densities that live on the ridge-line approximated by \autoref{eqn:Pareto}; \autoref{tab:SourceCounts} lists the same for the number of sources we could expect to detect using telescopes with different resolution and sensitivity thresholds.

\subsection{The case of M87 and the EHT} \label{sec:M87}

As of the writing of this paper, the SMBH in M87 is the only one whose shadow size ($\sim$40\,\uas) and horizon-scale flux density ($\sim$0.5\,Jy at 230\,GHz) have been directly imaged\footnote{The second shadow-resolved black hole that the EHT has targeted -- the Milky Way SMBH Sgr A* -- does not present a relevant comparison for this work because it is located in our own Galaxy, and it therefore does not fit within our modeling framework. In addition, Sgr~A* has an additional observing constraint beyond those given in \autoref{sec:PopSourceCounts}: it is heavily scattered by the ionized interstellar medium along its line of sight, so high-resolution observations must be conducted at correspondingly high frequencies of $\nu \gsim 1\,{\rm THz}/\sqrt{ \theta_r / 1\,\mu{\rm as}}$ \citep[e.g.,][]{Lo_1998,Bower_2006,Johnson_2018}. The scattering is significantly weaker for sources off the Galactic plane (such as M87), requiring only $\nu \gsim 30\,{\rm GHz}/\sqrt{ \theta_r / 1\,\mu{\rm as}}$ \citep[e.g.,][]{NE2001,Johnson_2015}. Thus, interstellar scattering is unlikely to significantly affect our estimates for observable source counts.} \citep{Paper1,Paper2,Paper3,Paper4,Paper5,Paper6}.  M87 thus presents a natural test case against which to compare our source counts predictions from \autoref{sec:PopSourceCounts}.  Our model predicts that the number of SMBHs having $\vartheta > 40$\,\uas and $S_{\nu} > 0.5$\,Jy should be between $\sim$0.03 and $\sim$0.23 for the lower and upper BHMF prescriptions, respectively.  Compared against the 1 object known to adhere to the chosen criteria, our model is systematically underpredicting the prevalence of M87. This underprediction may be explained at least in part if the local density of galaxies exceeds the cosmic mean, as suggested by, e.g., \cite{Dalya_2018}\footnote{We note that the overdensity in \cite{Dalya_2018} is driven almost entirely by the existence of the Virgo cluster, and there are other indications \citep[e.g.,][]{Tully_2019,Bohringer_2020} that when considering a somewhat larger volume (out to $\sim$100\,Mpc) the local Universe may actually be underdense.}, which violates our model assumption of a homogeneous distribution of SMBHs.  However, any such overdensity likely does not explain a discrepancy larger than a factor of $\sim$2, indicating that we may simply be finding ourselves on the high end of sampling variance.
We thus expect that using the existence and properties of M87 to extrapolate the number of SMBHs with smaller shadows or weaker flux densities will result in systematically over-optimistic predictions; i.e., more sources will be predicted than our modeling suggests the real Universe likely contains.

Similarly, the EHT is currently the only telescope to have successfully carried out shadow-resolved observations of a SMBH.  The number of sources that the EHT is able to resolve and detect the shadows for thus presents a test case against which to compare our source counts predictions from \autoref{sec:SourceCountsInt}.  The EHT currently relies on observing with ALMA as part of the array, and during the 2017 observing campaign that led to the published M87 black hole images, ALMA itself required in-beam sources with flux densities of ${\geq}0.5$\,Jy to perform the array phasing necessary for it to participate in VLBI observations \citep{Matthews_2018}.  For the purposes of estimating source counts, this phasing threshold effectively sets the sensitivity limit of the EHT.  In this case, our model predicts that for $\theta_r = 20$\,\uas and $\sigma_{\nu} = 0.5$\,Jy we should expect to resolve and detect up to $\sim$0.4 sources, similar to the projected number based on the above extrapolation using M87 as a benchmark.

However, the $0.5$\,Jy phasing threshold has since been relaxed by permitting the transfer of phase corrections to faint targets from nearby but bright out-of-beam calibrators, and even the on-source phasing threshold can potentially be lowered through refinement of the phasing algorithm.  Moving forward, the EHT may thus be able to observe much fainter targets.  In a best-case scenario in which the phasing threshold is reduced to mJy levels, these improvements could permit the nominal sensitivity of the EHT to be used for source count estimates.  Observing at 230~GHz, the EHT achieves $\theta_r \approx 20$\,\uas and $\sigma_{\nu} \approx 10^{-3}$\,Jy, for which our model predicts the number of accessible SMBHs to be between $\sim$0.6 and $\sim$5.7 for the lower and upper BHMF prescriptions, respectively. We thus predict that the EHT could potentially gain access to approximately an order of magnitude more shadow-resolved sources by improving its effective sensitivity to mJy levels in this way.

\subsection{Implications for array design} \label{sec:ArrayImplications}

More generally, the behavior of $N(\theta_r,\sigma_{\nu})$ -- in particular, the behavior of its gradient -- has implications for how an existing array can be most efficiently augmented to increase the number of accessible black hole shadows.  As mentioned in \autoref{sec:M87}, the EHT is currently operating with an angular resolution of $\theta_r \approx 20$\,\uas and an effective flux density sensitivity between $\sigma_{\nu} \approx 10^{-3}$\,Jy and $\sigma_{\nu} \approx 0.5$\,Jy.  This sensitivity range straddles the Pareto front for $\theta_r \approx 20$\,\uas (see \autoref{fig:pareto_front}), such that with $\sigma_{\nu} \approx 0.5$\,Jy the EHT array could most significantly increase the number of horizon-resolved black hole targets through improvements in sensitivity.  However, once the sensitivity improves beyond the Pareto threshold of ${\sim}70$\,mJy then the EHT will require enhanced angular resolution to increase the source counts further.  For instance, at a fixed sensitivity of $\sigma_{\nu} = 10^{-3}$\,Jy, an order-of-magnitude improvement in the angular resolution would yield an increase of roughly two orders of magnitude in the number of detectable black hole shadows; in contrast, while keeping the angular resolution fixed at $20$\,\uas, arbitrary improvements in sensitivity beyond $10^{-3}$\,Jy would not yield many additional sources.

In practice, an Earth-based array like the EHT is limited to a maximum physical baseline length of one Earth diameter, meaning that any significant angular resolution improvements must come from increasing the observing frequency.  A near-future aspiration for the EHT \citep{Paper2}, and a defining capability for the next-generation EHT \citep[ngEHT;][]{Doeleman_2019,Raymond_2021} will be to observe at a frequency of 345\,GHz.  At a fixed long-baseline sensitivity of $\sigma_{\nu} = 10^{-3}$\,Jy, we expect that the effective 50\% improvement in angular resolution over the current EHT should correspond to a factor of $\sim$3 increase in the number of detectable black hole shadows.  In contrast, at a fixed $\theta_r = 20$\,\uas, the doubling of the baseline sensitivity that the ngEHT is expected to provide will only increase the source counts by $\sim$10\%.

While angular resolution may ultimately limit the number of observable black hole shadows for ground-based interferometers like the EHT and ngEHT, sensitivity is expected to be the limiting factor for many prospective interferometers that network with space-based stations.  For instance, a baseline connecting a station on Earth to one located at the Earth-Sun L2 Lagrange point -- such as may be possible using the proposed Millimetron \citep{Kardashev_2014} or Origins \citep{Wiedner_2021} space telescopes -- would have a finest 230\,GHz angular resolution of $\theta_r \approx 0.2$\,\uas.  At this resolution, we expect that a sensitivity of $\sigma_{\nu} \lesssim 10^{-4}$\,Jy would be required to detect even a single object.  To achieve this sensitivity level, a 10-meter dish observing at 230\,GHz as part of a baseline with the phased ALMA array would require a time-bandwidth product of ${\sim}3 \times 10^{12}$ (e.g., 3 minutes of on-source integration time using 16\,GHz of bandwidth), which is already larger than achieved by the EHT.  Improving the sensitivity to $10^{-5}$\,Jy would require a time-bandwidth product that is two orders of magnitude larger still (e.g., 2 hours of on-source integration time using 32\,GHz of bandwidth), and pushing to $10^{-6}$\,Jy would require an additional two orders beyond that (e.g., 5 days of on-source integration time using 64\,GHz of bandwidth).  Achieving the ${\sim}10^{-6}$\,Jy Pareto front flux density corresponding to a ${\sim}0.2$\,\uas angular resolution thus imposes demanding sensitivity and stability requirements, and we expect that the number of sources accessible using long ($\gg$\,1 Earth diameter) Earth-space baselines will be sensitivity-limited rather than resolution-limited.

\section{Summary and conclusions} \label{sec:Summary}

Motivated by the success of the EHT and the promise of next-generation radio interferometric facilities, we have presented a framework for estimating the number of black hole shadows that are expected to be observationally accessible to different telescopes.  Given assumptions about the distribution of SMBHs across mass, accretion rate, and redshift, we use a semi-analytic ADAF-based SED model to derive estimates for the number of SMBHs with detectable and optically thin horizon-scale emission as a function of angular resolution, flux density sensitivity, and observing frequency.  Using a simple analytic prescription for the interferometric flux density distribution expected from black hole photon rings, we further decompose the SMBH source count estimates into the number of objects for which we could expect to observe first- and second-order photon rings.

Our main findings can be organized into two categories.  First, we provide the following characterizations of the SMBH population:

\begin{itemize}
    \item \autoref{fig:source_counts_230GHz_taucut} shows the distribution of observationally accessible SMBH shadows, predicting that large numbers (${>}10^6$ with ${\sim}0.1$\,\uas resolution and ${\sim}1\,\mu{\rm Jy}$ sensitivity) of objects should have resolvable horizon-scale emission at (sub)millimeter wavelengths.
    \item \autoref{fig:source_counts_230GHz_taucut_rescut} shows the angular resolution and sensitivity that an interferometer would require to observe the black hole shadows for this same population of SMBHs.
    \item For any particular choice of angular resolution and sensitivity, the population density of SMBHs with observable shadows generally increases toward higher redshifts and toward smaller black hole masses (see \autoref{fig:Mz_distributions_230GHz}).  As a consequence, a majority of observable shadows are expected to have angular sizes that fall close to the resolution limit.
    \item The bulk population of SMBHs with observable $n=1$ photon rings starts to become accessible at angular resolutions of ${\lesssim}2$\,\uas and flux density sensitivities of ${\lesssim}0.5$\,mJy (see \autoref{fig:source_counts_photon_ring_decomp} and \autoref{tab:SourceCounts}).  Similarly, the $n=2$ population is accessible for angular resolutions of ${\lesssim}0.1$\,\uas and flux density sensitivities of ${\lesssim}5$\,$\mu$Jy.
\end{itemize}

\noindent We also consider the implications of these findings for current and future interferometric facilities:

\begin{itemize}
    \item The current effective sensitivity of the EHT is insufficient to maximally utilize its angular resolution.  We predict that as many as ${\sim}$5 additional horizon-resolved sources could become accessible by improving the effective sensitivity of the EHT from $\sim$0.5\,Jy to $<$70\,mJy.  ALMA should be sufficiently sensitive to achieve phased observations on sources with flux densities at this level, so an important next step will be to identify the specific sources that then become accessible.
    \item Once the effective sensitivity of the EHT improves beyond the $\sim$tens of mJy level, a large (i.e., order-of-magnitude) additional increase in the number of observable black hole shadows can only be achieved by improving the angular resolution.  We predict that an ngEHT observing at 345\,GHz should have access to ${\sim}$3 times as many sources as the EHT observing at 230\,GHz.
    \item Future telescopes that observe with ${\lesssim}1$\,\uas angular resolution, such as could be achieved using Earth-space interferometry, will require flux density sensitivities of ${\ll}$1\,mJy to detect large numbers of black hole shadows.
\end{itemize}

In carrying out our analyses we have produced a library of synthetic SEDs and several tables of source counts\footnote{\url{http://dx.doi.org/10.17632/8pj73cy7vx.1}}, as well as the code used to generate each SED\footnote{\url{https://github.com/dpesce/LLAGNSED}}.  The source count tables provide the predicted number of black hole shadows, $n=1$ photon rings, and $n=2$ photon rings accessible using different combinations of angular resolution, flux density sensitivity, and frequency.  These resources may be useful for determining the specifications of future telescopes that aim to observe a large population of SMBH shadows or higher-order photon rings.  Once such observations have been carried out, the predictive framework developed in this paper could be inverted so that the source counts become inputs rather than outputs, in turn providing constraints on the distribution of SMBH masses and accretion rates across cosmic history.

\acknowledgments

We thank Gary Melnick for motivating conversations that sparked initial interest in pursuing this project.  We also thank Avery Broderick, Tim Davis, Jason Dexter, and the anonymous referee for constructive comments that improved the quality of the paper.  Support for this work was provided by the NSF through grants AST-1952099, AST-1935980, AST-1828513, AST-1440254, AST-1816420, and OISE 1743747, and by the Gordon and Betty Moore Foundation through grant GBMF-5278.  This work has been supported in part by the Black Hole Initiative at Harvard University, which is funded by grants from the John Templeton Foundation and the Gordon and Betty Moore Foundation to Harvard University.
NN acknowledges funding from Nucleo Milenio TITANs (NCN19$-$058).

\bibliography{references}{}
\bibliographystyle{aasjournal}

\clearpage
\appendix
\numberwithin{equation}{section}

\section{SED model} \label{app:SEDmodel}

We use a spectral energy distribution (SED) model for ADAFs that largely follows the formalism presented in \citetalias{Mahadevan_1997}, though we introduce a number of modifications that update the SED to align it better with more recent work.  In this section we detail these modifications, propagate them into the relevant expressions from \citetalias{Mahadevan_1997} and \citet[][hereafter \citetalias{Narayan_1995b}]{Narayan_1995b}, and describe the resulting SED model.

In determining the form of the SED, the primary equation we aim to solve is one of energy balance between the heating and cooling of the electrons in the flow. Following \citetalias{Mahadevan_1997} Eq.\,(8), we have

\begin{equation}
Q^{-} + Q^{\text{adv,e}} = Q^{\text{ie}} + \delta Q^{+} , \label{eqn:EnergyBalance}
\end{equation}

\noindent where $Q^{+}$ is the total viscous heating rate, $\delta$ is the fraction of this heating rate that goes directly to the electrons, $Q^{\text{ie}}$ is the rate of energy transfer from the ions to the electrons, $Q^{-}$ is the total radiative cooling rate of the electrons, and we have introduced an additional term $Q^{\text{adv,e}}$ that accounts for the electron energy that gets advected into the black hole. We note that in the extremely low accretion regime considered here, energy loss from neutrino cooling is negligible. The radiative cooling term is given by:

\begin{equation}
Q^{-} = P_{\text{synch}} + P_{\text{compt}} + P_{\text{brems}} , \label{eqn:TotalCooling}
\end{equation}

\noindent where $P_{\text{synch}}$, $P_{\text{compt}}$, and $P_{\text{brems}}$ correspond to the power emitted in synchtrotron, inverse Compton, and bremsstrahlung radiation, respectively.  It is the combined contributions from these three emission processes that ultimately constitute our model SED.

\begin{figure*}
    \centering
    \includegraphics[width=1.0\textwidth]{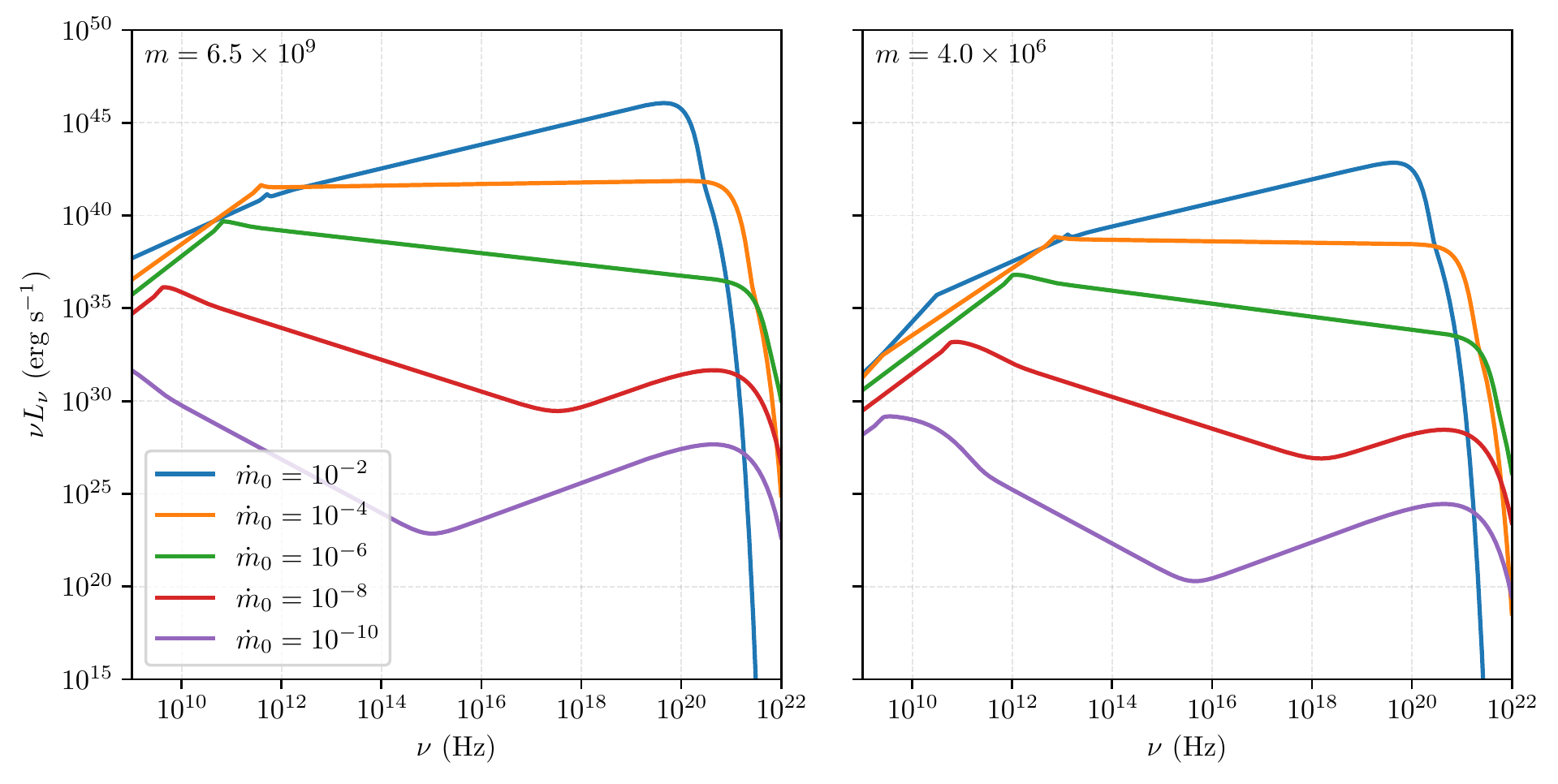}
    \caption{Example SEDs produced from the model described in \autoref{app:SEDmodel} plotted for a range of accretion rates.  The left panel shows SEDs for a SMBH mass similar to that of M87 ($6.5 \times 10^9$\,M$_{\odot}$; \citealt{Paper6}), and the right panel shows SEDs for a SMBH mass similar to that of Sgr A* ($4 \times 10^6$\,M$_{\odot}$; \citealt{Do_2019,Gravity_2019}).}
    \label{fig:SED_examples}
\end{figure*}

\begin{figure*}
    \centering
    \includegraphics[width=0.49\textwidth]{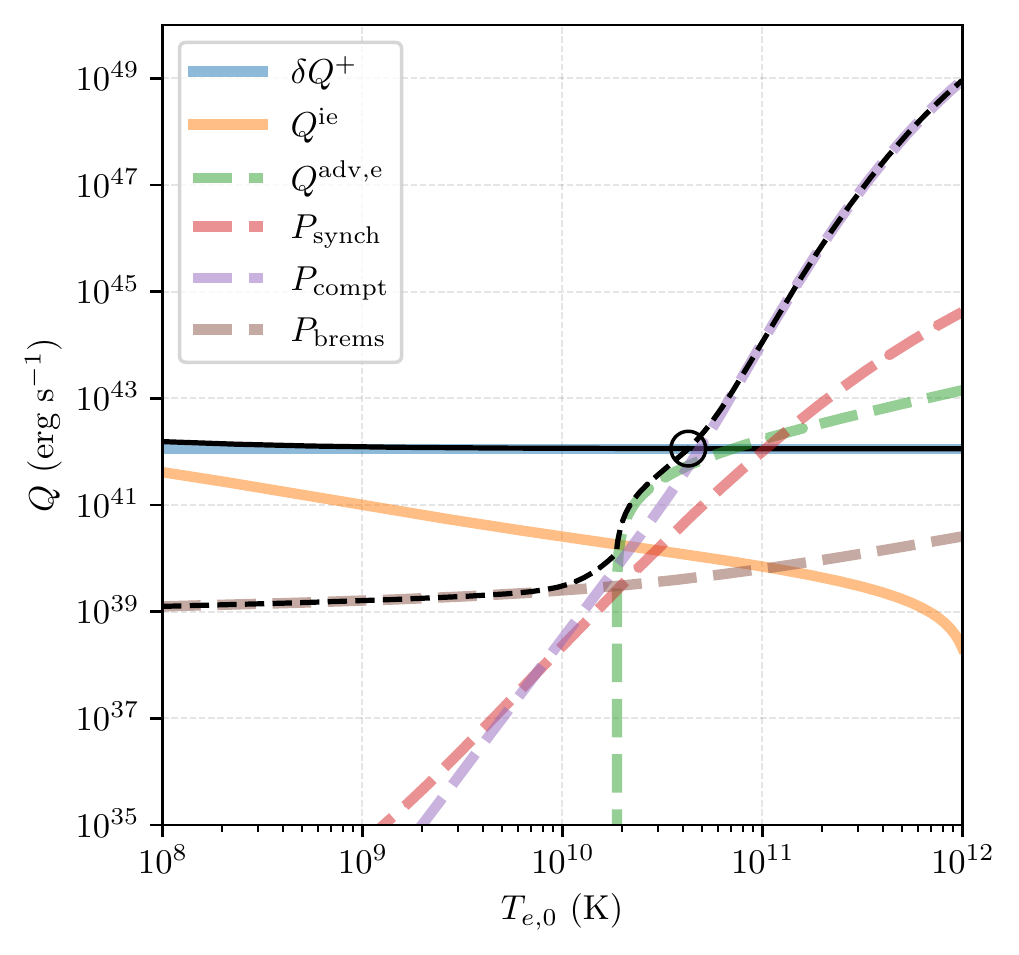}
    \includegraphics[width=0.49\textwidth]{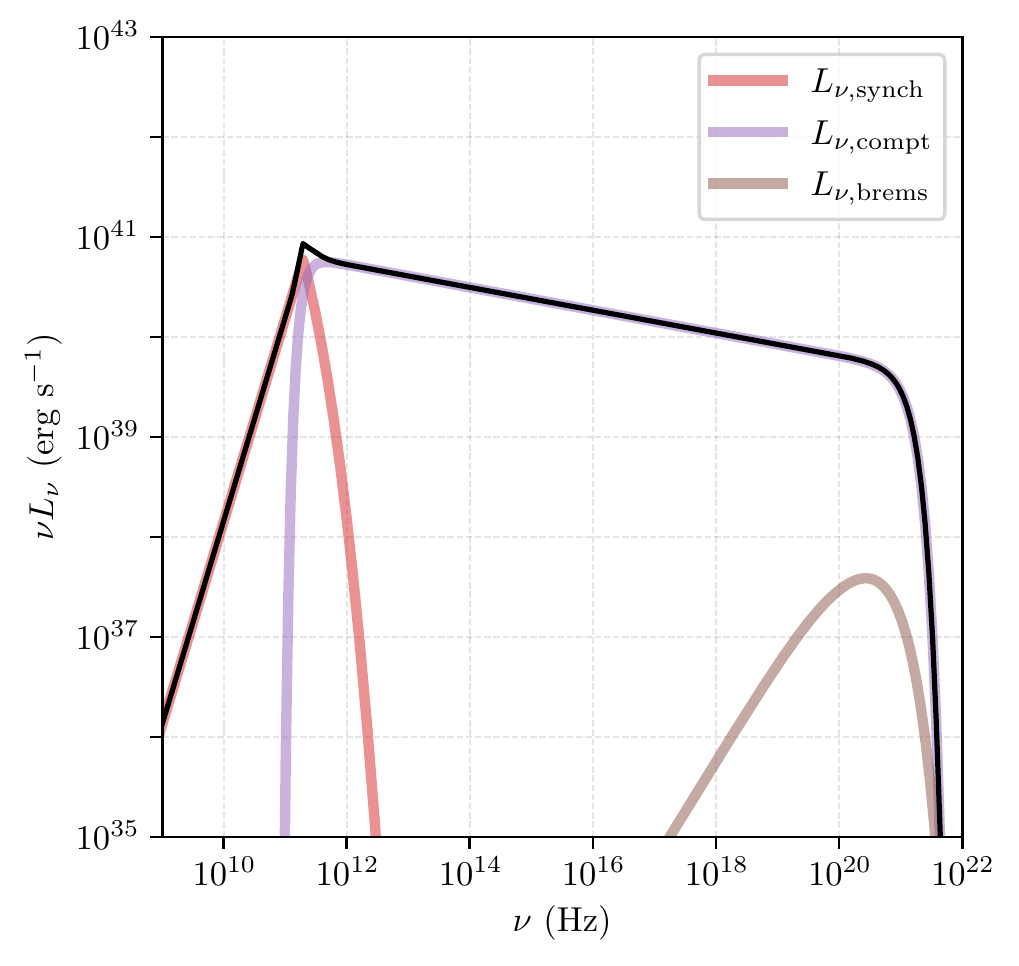}
    \caption{\textit{Left}: An example set of electron heating and cooling curves used in the SED model as a function of $T_{e,0}$, for an M87-like SMBH with $m = 6.5 \times 10^9$ and $\dot{m}_0 = 10^{-5}$.  Solid colored curves indicate sources of electron heating, while dashed colored curves indicate sources of electron cooling; the total heating and cooling are plotted as the black solid and dashed lines, respectively.  The intersection of these lines is circled and indicates where the heating and cooling are balanced (see \autoref{eqn:EnergyBalance}), which for this system occurs at $T_{e,0} = 4.3 \times 10^{10}$\,K. \textit{Right}: The SED corresponding to the solution from the left panel, with the individual contributions from synchrotron, inverse Compton, and bremsstrahlung emission plotted as colored curves and the combined spectrum plotted in black.  For a 17\,Mpc assumed distance to M87, the predicted 230\,GHz flux density is $\sim$1\,Jy.  This prediction agrees well with the horizon-scale flux density measured by the Event Horizon Telescope \citep{Paper4}.}
    \label{fig:SED_temperature}
\end{figure*}

The emission processes of interest for this paper depend on the electron temperature, which is determined self-consistently such that the total heating and cooling satisfy \autoref{eqn:EnergyBalance}.  The left panel of \autoref{fig:SED_temperature} shows the various model contributions to the electron heating and cooling as a function of electron temperature for an example M87-like system, and the right panel shows the corresponding predicted SED as a function of frequency.  The right panel of \autoref{fig:nup_and_Te_ve_m_mdot} shows the derived temperatures as a function of $m$ and $\dot{m}$.  \autoref{tab:SED_model_params} provides a list of the various parameters used in the SED model, and \autoref{fig:SED_examples} shows example SEDs.

\subsection{Flow equations}

We take the underlying accretion flow properties to be described by the self-similar models developed by \citetalias{Narayan_1995b}, in which the relevant parameters are the black hole mass $M$, the accretion rate $\dot{M}$, the radius $R$, the viscosity parameter $\alpha$, the ratio of gas to magnetic pressure $\beta$,\footnote{We note that our definition for $\beta$ differs from that used in \citetalias{Mahadevan_1997}; \citetalias{Mahadevan_1997} uses the ratio of the gas pressure to the total pressure, while we use the ``plasma beta'' convention (i.e., ratio of gas pressure to magnetic pressure).  If we denote the \citetalias{Mahadevan_1997} parameter as $\beta_{\text{M97}}$, then the two are related by $\beta_{\text{M97}} = \beta / (1 + \beta)$.} and the fraction $f$ of viscously dissipated energy that gets advected into the black hole.  Following \citetalias{Mahadevan_1997}, we use scaled quantities,

\begin{subequations}
\begin{equation}
M = \left( 1 \text{ \msun} \right) m ,
\end{equation}
\begin{eqnarray}
R & = & r R_S \nonumber \\
& = & \left( 2.953 \times 10^5 \text{ cm} \right) m\,r ,
\end{eqnarray}
\begin{eqnarray}
\dot{M} & = & \dot{m} \dot{M}_{\text{Edd}} \nonumber \\
& = & \left( \dot{m}_0 \dot{M}_{\text{Edd}} \right) \left( \frac{R}{R_S} \right)^s \nonumber \\
& = & \left( 1.399 \times 10^{18} \text{ g s}^{-1} \right) m \, \dot{m}_0 \, r^s , \label{eqn:MdotVsRadius}
\end{eqnarray}
\end{subequations}

\noindent where $R_S = 2 G M / c^2$ is the Schwarzschild radius, $\dot{M}_{\text{Edd}} = L_{\text{Edd}} / \eta c^2$ is the Eddington accretion rate, and we have taken the radiative efficiency $\eta$ to be 0.1.  Here, the difference between \autoref{eqn:MdotVsRadius} and  \citetalias{Mahadevan_1997} Eq.\,(4) comes from our adoption of the radius-dependent accretion rate from \cite{Blandford_1999}, which accounts for outflowing material via a radial dependence of the mass accretion rate with power-law index $s$.

The self-similar equations describing the accretion flow, \citetalias{Mahadevan_1997} Eq.\,(5), become

\begin{subequations}
\begin{eqnarray}
\rho & = & \frac{\dot{M}}{4 \pi H \alpha c_1 \sqrt{G M R}} \nonumber \\
& = & \left( 6.022 \times 10^{-5} \text{ g cm}^{-3} \right) \alpha^{-1}\,c_1^{-1}\,m^{-1}\, \dot{m}_0\,r^{-(3/2) + s} ,
\end{eqnarray}
\begin{eqnarray}
n_e & = & \frac{\rho}{\mu_e m_p} \nonumber \\
& = & \left( 3.158 \times 10^{19} \text{ cm}^{-3} \right) \alpha^{-1}\,c_1^{-1}\,m^{-1}\, \dot{m}_0\,r^{-(3/2) + s} , \label{eqn:NumberDensityProfile}
\end{eqnarray}
\begin{eqnarray}
B & = & \sqrt{\frac{24 \pi c_3 G M \rho}{(1 + \beta) R}} \nonumber \\
& = & \left( 1.428 \times 10^9 \text{ G} \right) \left( 1 + \beta \right)^{-1/2} \alpha^{-1/2}\,c_1^{-1/2}\,c_3^{1/2}\,m^{-1/2}\,\dot{m}_0^{1/2}\,r^{-(5/4) + (s/2)} ,
\end{eqnarray}
\end{subequations}

\noindent where $\rho$ is the mass density, $B$ is the magnetic field strength, $n_e$ is the number density of electrons, $\alpha$ is the disk viscosity parameter \citep{Shakura_1973}, $H$ is the disk scale height (we have followed \citetalias{Mahadevan_1997} in setting $H = R$), $\mu_e = 1.14$ is the mean molecular weight \citepalias{Narayan_1995b}, and $c_1 \approx 0.5$ and $c_3 \approx 0.3$ are constants defined in \citetalias{Narayan_1995b} and specified in \autoref{tab:SED_model_params}.

We adopt a power-law radial profile for the electron temperature $T_e$ of the form

\begin{equation}
T_e = \frac{T_{e,0}}{r^{1-t}} , \label{eqn:TemperatureProfile}
\end{equation}

\noindent with $t \leq 1$.  From \citetalias{Narayan_1995b} Eq.\,(2.16), the two-temperature accretion flow must satisfy

\begin{equation}
T_i + 1.08 T_e = \left( 6.66 \times 10^{12} \text{ K} \right) \left( 1 + \beta \right)^{-1} \beta\,c_3\,r^{-1} ,
\end{equation}

\noindent where $T_i$ is the ion temperature.  Setting $T_i = T_e$ at some maximum radius $r = r_{\text{max}}$ yields an expression for $t$,

\begin{equation}
t = \frac{1}{\ln\left( r_{\text{max}}\right)} \ln\left( \frac{\left( 6.66 \times 10^{12} \text{ K} \right) \beta c_3}{2.08 T_{e,0} \left( 1 + \beta \right)} \right) ,
\end{equation}

\noindent such that $T_i > T_e$ for all $r < r_{\text{max}}$.

\subsection{Heating}

The plasma in an ADAF is heated by viscous forces, with the total heating rate per unit volume denoted as $q^+$.  Some fraction $\delta$ of this energy gets deposited into the electrons, while the remaining fraction $(1-\delta)$ heats the ions.  The ions can transfer thermal energy to the electrons via Coulomb collisions, with the rate of this transfer denoted by $q^{\text{ie}}$, and the electrons can radiate energy away at a rate $q^-$.  Taken altogether, energy balance yields advected energy rates of

\begin{subequations}
\begin{equation}
q^{\text{adv,i}} = (1 - \delta) q^+ - q^{\text{ie}} ,
\end{equation}
\begin{equation}
q^{\text{adv,e}} = \delta q^+ + q^{\text{ie}} - q^- ,
\end{equation}
\end{subequations}

\noindent for the ions ($q^{\text{adv,i}}$) and electrons ($q^{\text{adv,e}}$).  The ion heating is driven by viscous dissipation, while the dominant electron heating source depends on the accretion rate; at high accretion rates the ion-electron heating is dominant, whereas at low accretion rates the viscous heating is more important.

\citetalias{Narayan_1995b} give an expression for the viscous heating rate per unit volume,

\begin{eqnarray}
q^+ & = & \frac{3 \rho \alpha c_1 c_3}{2 R f (1 + \beta)} \left( \frac{G M}{R} \right)^{3/2} \nonumber \\
& = & \left( 2.914 \times 10^{21} \text{ erg cm}^{-3} \text{ s}^{-1} \right) f^{-1} (1 + \beta)^{-1} c_3^{1/2} m^{-2} \dot{m}_0 r^{-4+s} ,
\end{eqnarray}

\noindent where $f$ is the fraction of viscously dissipated energy that gets advected into the black hole.  The total (i.e., volume-integrated) viscous heating rate is then given by

\begin{equation}
Q^+ = \left( 9.430 \times 10^{38} \text{ erg s}^{-1} \right) f^{-1} (1 + \beta)^{-1} c_3^{1/2} m \dot{m}_0 \times \begin{cases}
(1 - s)^{-1} \left( r_{\text{min}}^{-1 + s} - r_{\text{max}}^{-1 + s} \right) , & s \neq 1 \\
\ln\left( r_{\text{max}} / r_{\text{min}} \right) , & s = 1
\end{cases} , \label{eqn:Qplus}
\end{equation}

\noindent where $r_{\text{min}}$ and $r_{\text{max}}$ are the minimum and maximum radius, respectively.\footnote{We note and correct an error in the original expression for $Q^+$ from \citetalias{Mahadevan_1997} Eq.\,(9), for which the exponent of the $c_3$ term should be $1/2$ rather than 1.}

The heating rate per unit volume of the electrons from Coulomb interactions with protons is given by \cite{Stepney_1983},

\begin{eqnarray}
q^{\text{ie}} & = & \frac{3 m_e}{2 m_p} n_e n_i \sigma_{\text{T}} c \ln(\Lambda) \left( \frac{k (T_i - T_e)}{K_2(1/\theta_e) K_2(1/\theta_i)} \right) \left[ \frac{2 (\theta_e + \theta_i)^2 + 1}{\theta_e + \theta_i} K_1\left( \frac{\theta_e + \theta_i}{\theta_e \theta_i} \right) + 2 K_0\left( \frac{\theta_e + \theta_i}{\theta_e \theta_i} \right) \right] \nonumber \\
& \approx & \left( 5.624 \times 10^{-32} \text{ erg cm}^{3} \text{ s}^{-1} \text{ K}^{-1} \right) \frac{n_e^2 (T_i - T_e)}{K_2(1/\theta_e)} \left( 2 + 2\theta_e + \frac{1}{\theta_e} \right) e^{-1/\theta_e} , \label{eqn:IonElectronHeating}
\end{eqnarray}

\noindent where $\theta_e = k T_e / m_e c^2$ is the dimensionless electron temperature, $\theta_i = k T_i / m_p c^2$ is the dimensionless ion temperature, $\ln(\Lambda) \approx 20$ is a Coulomb logarithm, $K_n$ represents a modified Bessel function of the $n$th order, and we have assumed $n_e = n_i$.  In the second line we have adopted the approximation from \citetalias{Mahadevan_1997}.\footnote{We note and correct an error in the original expression for $q^{\text{ie}}$ from \citetalias{Mahadevan_1997} Eq.\,(10), for which the exponent of the $r$ term should be $-3$ rather than $-1$.}  We note that in evaluating the prefactor in \autoref{eqn:IonElectronHeating} we have followed \citetalias{Narayan_1995b} and multiplied by an additional factor of 1.25 to account for the ions containing a mixture of roughly 75\% hydrogen and 25\% helium.

The volume-integrated ion-electron heating rate is given by

\begin{equation}
Q^{\text{ie}} = \left( 3.236 \times 10^{17} \text{ cm}^3 \right) m^3 \int_{r_{\text{min}}}^{r_{\text{max}}} q^{\text{ie}} r^2 \, \deriv r ,
\end{equation}

\noindent which does not have an analytic form and so must be integrated numerically.

\subsection{Cooling}

The observed emission in radio and (sub)millimeter bands is dominated by synchrotron radiation, but the primary electron cooling mechanisms also include bremsstrahlung and inverse Compton radiation.  Each of these emission mechanisms contributes to $q^-$, and each depends on the electron temperature $T_e$.

\subsubsection{Synchrotron emission}

We use a form for the synchrotron spectrum from \cite[][see also \citetalias{Narayan_1995b,Mahadevan_1997}]{Mahadevan_1996}, which assumes an isotropic distribution of relativistic electrons.  The synchrotron spectral emissivity is given by

\begin{equation}
\epsilon_{\text{synch},\nu} = \left( 4.43 \times 10^{-30} \text{ erg s}^{-1} \text{ Hz}^{-2} \right) \frac{4 \pi n_e \nu}{K_2(1/\theta_e)} M(x_M) , \label{eqn:SynchrotronEmissivity}
\end{equation}

\noindent where we assume the relativistic limit for $M(x_M)$,

\begin{equation}
M(x_M) = \frac{4.0505}{x_M^{1/6}} \left( 1 + \frac{0.40}{x_M^{1/4}} + \frac{0.5316}{x_M^{1/2}} \right) \exp\left( -1.8899 x_M^{1/3} \right) ,
\end{equation}

\noindent and $x_M$ is a dimensionless frequency,

\begin{subequations}
\begin{equation}
x_M = \frac{2 \nu}{3 \nu_b \theta_e^2} , \label{eqn:DimensionlessSynchFreq}
\end{equation}
\begin{eqnarray}
\nu_b & = & \frac{e B}{2 \pi m_e c} \nonumber \\
& = & \left( 3.998 \times 10^{15} \text{ Hz} \right) \left( 1 + \beta \right)^{-1/2} \alpha^{-1/2} c_1^{-1/2} c_3^{1/2} m^{-1/2} \dot{m}_0^{1/2} r^{-(5/4) + (s/2)} .
\end{eqnarray}
\end{subequations}

\noindent \autoref{eqn:SynchrotronEmissivity} assumes optically thin emission, but below some critical frequency $\nu_c$ (which is a function of radius) we expect the synchrotron to be optically thick and thus described by a blackbody spectrum.  We follow \citetalias{Mahadevan_1997} and determine $\nu_c(r)$ by equating emission within a volume of radius $r$ to the Rayleigh-Jeans blackbody emission from a spherical surface at that radius,

\begin{equation}
\epsilon_{\text{synch},\nu_c} \left( \frac{4 \pi R^3}{3} \right) = 4 \pi R^2 \left( \frac{2 \pi \nu_c^2 k T_e}{c^2} \right) , \label{eqn:CriticalSynchFreq}
\end{equation}

\noindent which lends itself to a prescription for estimating the optical depth more generally of

\begin{eqnarray}
\tau & = & \frac{\epsilon_{\text{synch},\nu} R c^2}{6 \pi \nu^2 k T_e} \nonumber \\
& = & \left( \frac{\nu}{\nu_c} \right)^{-2} \left( \frac{\epsilon_{\text{synch},\nu_c}}{\epsilon_{\text{synch},\nu}} \right) . \label{eqn:OpticalDepth}
\end{eqnarray}

\begin{figure}
    \centering
    \includegraphics[width=1.00\textwidth]{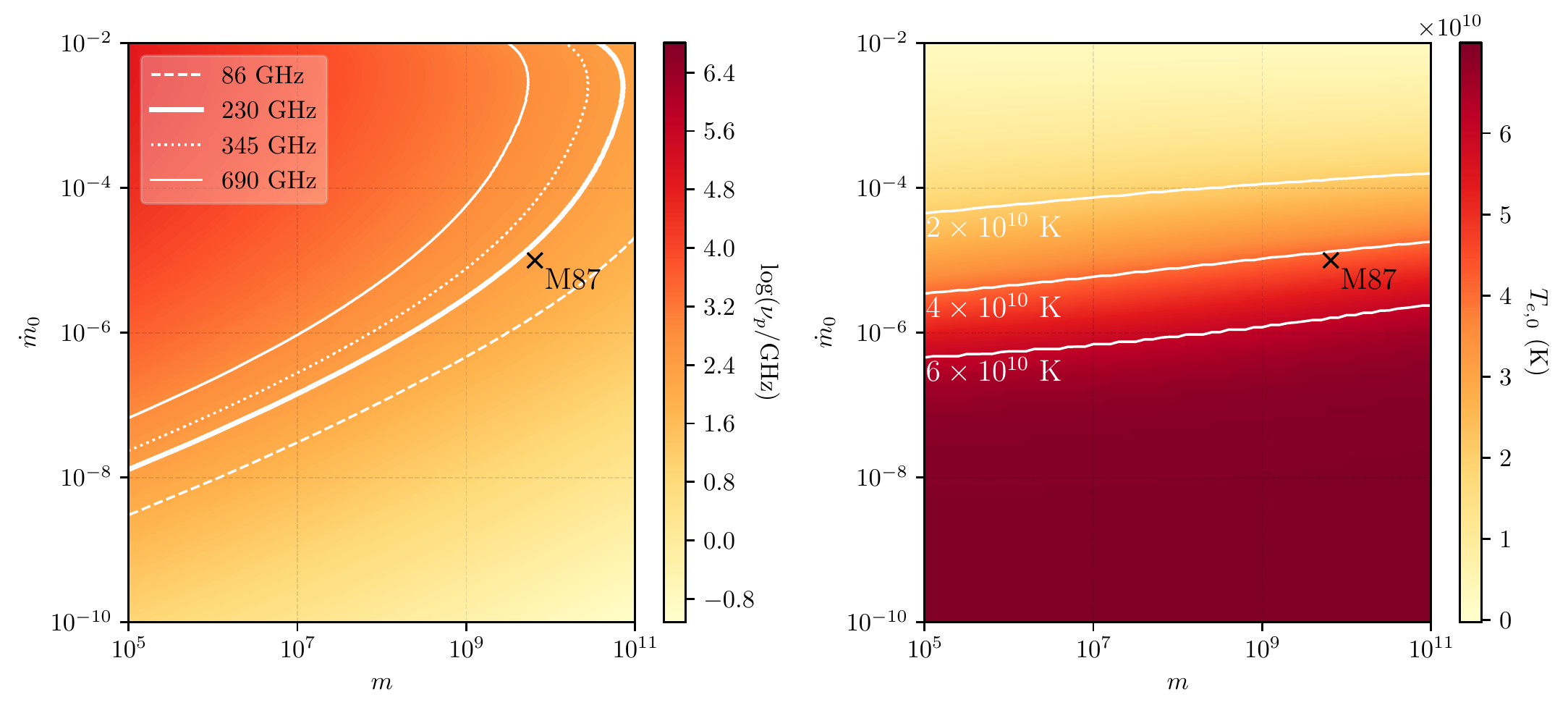}
    \caption{\textit{Left}: The peak synchrotron frequency, $\nu_p$, at $R_{\text{min}}$ as a function of $m$ and $\dot{m}$.  The white curves show contours at four different observing frequencies of interest, such that at any given frequency objects living to the lower right of the curve are expected to have optically thin emission.  \textit{Right}: The self-consistent value determined for $T_{e,0}$ as a function of $m$ and $\dot{m}$.  In both panels, the approximate mass ($6.5 \times 10^9$\,\msun) and accretion rate ($\lambda = 10^{-5}$) corresponding to the SMBH in M87 is marked in black.}
    \label{fig:nup_and_Te_ve_m_mdot}
\end{figure}

\noindent We numerically solve \autoref{eqn:CriticalSynchFreq} for $\nu_c$ at $R_{\text{min}}$ and $R_{\text{max}}$, yielding a peak frequency $\nu_p$ at $R_{\text{min}}$ (with luminosity $L_p$) and a minimum frequency $\nu_m$ at $R_{\text{max}}$ (with luminosity $L_m$); the left panel of \autoref{fig:nup_and_Te_ve_m_mdot} shows how $\nu_p$ changes with $m$ and $\dot{m}$.  We take the synchrotron spectrum to be blackbody (i.e., optically thick) at frequencies below $\nu_m$, optically thin with an emissivity described by \autoref{eqn:SynchrotronEmissivity} at frequencies above $\nu_p$, and a power law at intermediate frequencies.  That is,

\begin{equation}
L_{\nu,\text{synch}} = \begin{cases}
\left( 1.058 \times 10^{-24} \text{ erg s}^{-1} \text{ Hz}^{-3} \text{ K}^{-1} \right) T_{e,0} m^2 \nu^2 r_{\text{max}}^{1+t} , & \nu < \nu_m \\
L_m \left( \frac{\nu}{\nu_m} \right)^{\ln(L_p/L_m) / \ln(\nu_p/\nu_m)} , & \nu_m \leq \nu \leq \nu_p \\
\left( 1.896 \times 10^{8} \text{ erg s}^{-1} \text{ Hz}^{-2} \right) \frac{M(x_M)}{K_2(1/\theta_e)} \alpha^{-1} c_1^{-1} m^2 \dot{m}_0 \nu r_{\text{min}}^{(3/2) + s} , & \nu > \nu_p
\end{cases} .
\end{equation}

\noindent The total emitted synchrotron power is then the integral of $L_{\nu,\text{synch}}$ over frequency,

\begin{equation}
P_{\text{synch}} = \int_0^{\infty} L_{\nu,\text{synch}} \, \deriv \nu ,
\end{equation}

\noindent which we evaluate numerically.

\subsubsection{Bremsstrahlung emission}

We use an expression for the bremsstrahlung emission that follows \citetalias{Mahadevan_1997} Eq.\,(27),

\begin{eqnarray}
q_{\text{brems}} & = & \left( 1.48 \times 10^{-22} \text{ erg cm}^{3} \text{ s}^{-1} \right) n_e^2 F(\theta_e) \nonumber \\
& = & \left( 1.476 \times 10^{17} \text{ erg cm}^{-3} \text{ s}^{-1} \right) \alpha^{-2} c_1^{-2} m^{-2} \dot{m}_0^2 F(\theta_e) r^{-3 + 2 s} ,
\end{eqnarray}

\noindent where

\begin{equation}
F(\theta_e) = \begin{cases}
4 \left( \frac{2 \theta_e}{\pi^3} \right)^{1/2} \left( 1 + 1.781 \theta_e^{1.34} \right) + 1.73 \theta_e^{3/2} \left( 1 + 1.1 \theta_e + \theta_e^2 - 1.25 \theta_e^{5/2} \right) , & \theta_e \leq 1 \\
\left( \frac{9 \theta_e}{2 \pi} \right) \left[ \ln\left( 1.123 \theta_e + 0.48 \right) + 1.5 \right] + 2.30 \theta_e \left[ \ln\left( 1.123 \theta_e \right) + 1.28 \right] , & \theta_e > 1
\end{cases} .
\end{equation}

\noindent The volume-integrated power emitted in bremsstrahlung radiation will then be

\begin{eqnarray}
P_{\text{brems}} & = & \left( 3.236 \times 10^{17} \text{ cm}^3 \right) m^3 \int_{r_{\text{min}}}^{r_{\text{max}}} q_{\text{brems}} r^2 \, \deriv r \nonumber \\
& = & \left( 4.776 \times 10^{34} \text{ erg s}^{-1} \right) \alpha^{-2} c_1^{-2} m \dot{m}_0^2 \int_{r_{\text{min}}}^{r_{\text{max}}} F(\theta_e) r^{-1 + 2 s} \, \deriv r ,
\end{eqnarray}

\noindent with a spectral dependence given by

\begin{eqnarray}
L_{\nu,\text{brems}} & = & \left( 2.292 \times 10^{24} \text{ erg s}^{-1} \text{ Hz}^{-1} \text{ K} \right) \alpha^{-2} c_1^{-2} m \dot{m}_0^2 T_{e,0}^{-1} \int_{r_{\text{min}}}^{r_{\text{max}}} F(\theta_e) \exp\left( - \frac{h \nu}{k T_e} \right) r^{-2 + 2 s + t} \, \deriv r .
\end{eqnarray}

\noindent We integrate both of the above expressions numerically.

\subsubsection{Inverse Compton emission}

We follow \citetalias{Mahadevan_1997} in considering Comptonization only of synchrotron photons emitted predominantly at the peak frequency $\nu_p$, for which the spectrum in the temperature range of interest is expected to be a power law,

\begin{equation}
L_{\nu,\text{compt}} = L_p \left( \frac{\nu}{\nu_p} \right)^{-\alpha_c} .
\end{equation}

\noindent The power-law index $\alpha_c$ is determined by both how frequently photons are scattered (which is determined by the optical depth of the scattering process) and by how much a photon gets amplified during a typical scattering event.  We use an expression for the optical depth to electron scattering $\tau_{\text{es}}$ adapted from \citetalias{Narayan_1995b} Eq.\,(2.15),

\begin{eqnarray}
\tau_{\text{es}} & = & 2 n_e \sigma_T R_{\text{min}} \nonumber \\
& = & 6.205 \alpha^{-1} c_1^{-1} \dot{m}_0 r_{\text{min}}^{-(1/2) + s} .
\end{eqnarray}

\noindent We take the mean amplification factor $A$ from \citetalias{Mahadevan_1997} Eq.\,(32) (originally inspired from \citealt{Rybicki_Lightman}),

\begin{equation}
A = 1 + 4 \theta_{e,0} + 16 \theta_{e,0}^2 ,
\end{equation}

\noindent which together with $\tau_{\text{es}}$ determines the power-law slope for the Compton emission,

\begin{equation}
\alpha_c = - \frac{\ln\left( \tau_{\text{es}} \right)}{\ln(A)} .
\end{equation}

\noindent The total Compton power will then be given by the integral of $L_{\nu,\text{compt}}$ up to the maximum final frequency of a Comptonized photon ($\nu_f = 3 k T_{e,0} / h$),

\begin{eqnarray}
P_{\text{compt}} & = & \int_{\nu_p}^{\nu_f} L_{\nu,\text{compt}} \, \deriv \nu \nonumber \\
& = & \frac{\nu_p L_p}{1 - \alpha_c} \left[ \left( \frac{\nu_f}{\nu_p} \right)^{1-\alpha_c} - 1 \right] .
\end{eqnarray}

\subsubsection{Electron advection}

The \citetalias{Mahadevan_1997} model assumed that $T_e \ll T_i$, and it therefore ignored electron energy advection.  This assumption was reasonable for the parameters considered in that paper, particularly the choice of $\delta = m_e/m_p$. However, the modern view is that $\delta$ is much larger ($\approx 0.3$; see \citealt{Yuan_2014}). Such large values of $\delta$ make electrons significantly hotter, especially at very low $\dot{m}$, and so energy advection in electrons can no longer be ignored.

When electron advection is included, \citetalias{Mahadevan_1997} Eq.\,(8) gains an additonal term $Q^{\text{adv,e}}$ and becomes \autoref{eqn:EnergyBalance}. This advective cooling term is given by

\begin{equation}
Q^{\text{adv,e}} = \int_{R_{\text{min}}}^{R_{\text{max}}} 4\pi R^2 \left( n_e v
T_e \frac{ds_e}{dR}\right) \, \deriv R,
\end{equation}

\noindent where $s_e$ is the entropy per electron. Let us write

\begin{equation}
T_e \deriv s_e = \deriv u_e + p_e \deriv \left(\frac{1}{n_e}\right)
= \frac{k \deriv T_e}{\gamma_{\text{CV}} - 1} - \frac{kT_e}{n_e} \deriv n_e,
\end{equation}

\noindent where, following the approach described in \cite{Narayan_1994}, we express the specific heat at constant volume $C_V$ in terms of an effective $\gamma_{\text{CV}}$.  Substituting in \autoref{eqn:NumberDensityProfile} and \autoref{eqn:TemperatureProfile} and differentiating with respect to $R$ yields

\begin{equation}
T_e \frac{ds_e}{dR} = \frac{k T_{e,0}}{R_S r^{2-t}} \left( \frac{3}{2} - s - \frac{1-t}{\gamma_{\text{CV}} - 1} \right) .
\end{equation}

\noindent \cite{Sadowski_2017} provide an accurate fitting function for $\gamma_{\text{CV}}$, which we write as

\begin{equation}
\gamma_{\text{CV}} = \frac{20 \left( 2 + 8 \theta_e + 5 \theta_e^2 \right)}{3 \left( 8 + 40 \theta_e + 25 \theta_e^2 \right)} .
\end{equation}

\noindent Noting further that $n_e = \rho/\mu_e m_p$ and $\dot{M} = -4\pi R^2 v \rho$, we finally obtain

\begin{equation}
Q^{\text{adv,e}} = \left( 1.013 \times 10^{26} \text{ erg s}^{-1} \text{ K}^{-1} \right) m \dot{m}_0 T_{e,0} \int_{r_{\text{min}}}^{r_{\text{max}}} \left( \frac{1-t}{\gamma_{\text{CV}} - 1} - \frac{3}{2} + s \right) r^{s + t - 2} \, \deriv r , \label{eqn:ElectronAdvection}
\end{equation}

\noindent which we integrate numerically.  We note that there are conditions under which \autoref{eqn:ElectronAdvection} can yield a negative value for $Q^{\text{adv,e}}$; in these cases, we impose $Q^{\text{adv,e}} = 0$.

\subsection{Maximum mass accretion rate} \label{sec:CriticalMdot}

The ADAF solution ceases to exist above some critical mass accretion rate, $\dot{m}_{\text{crit}}$, where the accretion flow is no longer advection-dominated \citepalias{Narayan_1995b,Mahadevan_1997}.  Within the context of our SED model, this condition manifests as a maximum accretion rate above which there is no equilibrium temperature (i.e., the heating and cooling curves never cross).  We numerically determine a value of $\dot{m}_{\text{crit}} \approx 10^{-1.7}$, and so in this paper we only work with values of $\dot{m}_0 \leq 10^{-2}$.

\begin{figure}
    \centering
    \includegraphics[width=0.50\textwidth]{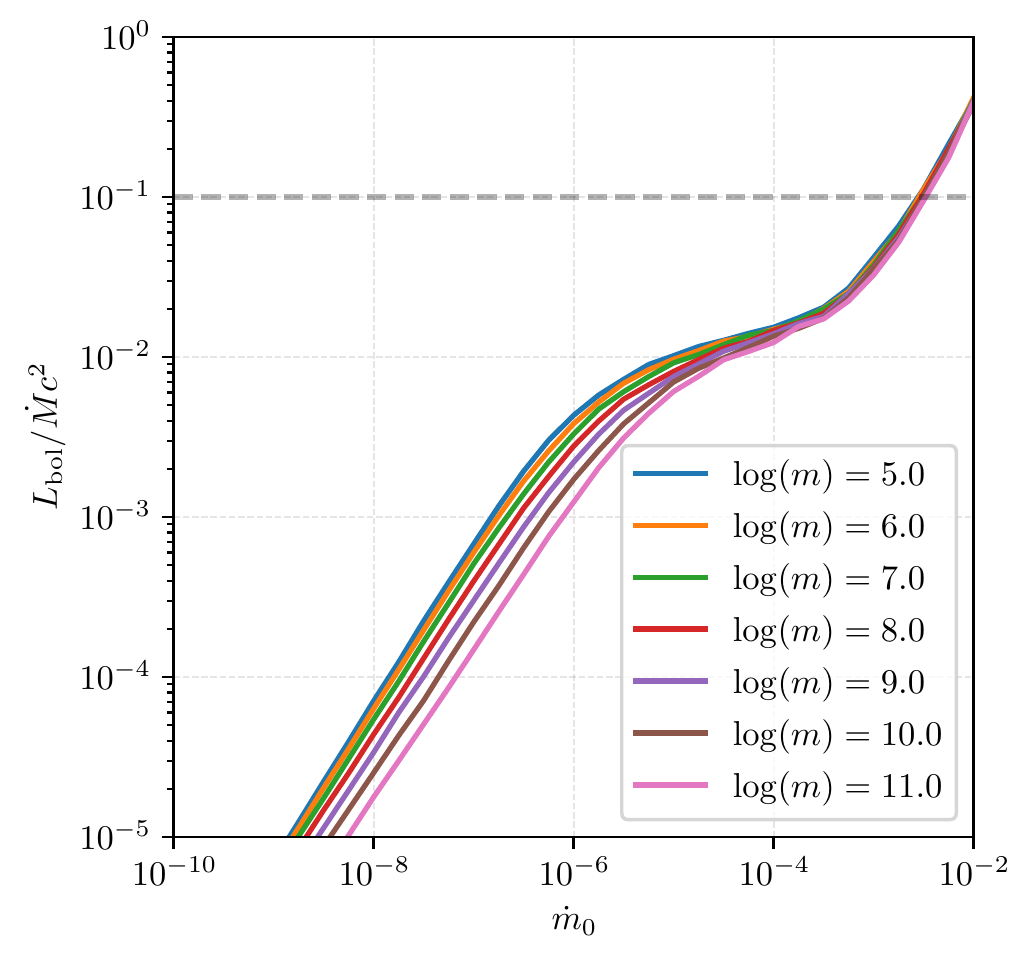}
    \caption{The ratio of the bolometric luminosity $L_{\text{bol}}$ to the accretion luminosity $\dot{M} c^2$ versus the accretion rate $\dot{m}$ and colored by SMBH mass $m$.  The dashed gray line indicates the input radiative efficiency of $\eta = 0.1$; for $\dot{m}_0 \gtrsim 10^{-2.5}$ the radiative efficiency of our model exceeds this input assumption.}
    \label{fig:luminosity_ratio_vs_lambda}
\end{figure}

In addition to the critical $\dot{m}$ above which no ADAF solutions exist, there is also a softer threshold accretion rate above which solutions do exist but our assumed input radiative efficiency of $\eta = 0.1$ is no longer consistent with the output of the SED model.  \autoref{fig:luminosity_ratio_vs_lambda} shows the predicted radiative efficiency from the model as a function of $\dot{m}_0$; we take the model radiative efficiency to be the ratio of the bolometric luminosity,

\begin{equation}
L_{\text{bol}} = \int_0^{\infty} \left( L_{\nu,\text{synch}} + L_{\nu,\text{compt}} + L_{\nu,\text{brems}} \right) \, \deriv \nu ,
\end{equation}

\noindent to the accretion rate equivalent luminosity, $\dot{M} c^2$.  Regardless of the input mass, the output radiative efficiency exceeds the assumed input value for $\dot{m}_0 \gtrsim 10^{-2.5}$.  Though this inconsistency reflects a physical limitation of the model, we note that given the ERDF prescription used in this paper (see \autoref{sec:ERDF}) it impacts only a small fraction of SMBHs (${<}$1\% for most $M$ and $z$, reaching a peak of ${\sim}$5\% for $M > 10^9$\,\msun and $z > 5$).

\begin{deluxetable}{Lcc}
\tablecolumns{3}
\tablewidth{0pt}
\tablecaption{SED model parameters\label{tab:SED_model_params}}
\tablehead{\colhead{\textbf{Parameter}} & \colhead{\textbf{Description}} & \colhead{\textbf{Default value}}}
\startdata
m & black hole mass in units of the solar mass; $m \equiv M/\text{\msun}$ & \ldots \\
\dot{m}_0 & mass accretion rate onto the black hole, in units of Eddington & \ldots \\
\eta & radiative efficiency & 0.1 \\
\beta & plasma beta; ratio of gas pressure to magnetic pressure &  10 \\
\alpha & viscosity parameter &  0.2 \\
s & power-law index for the mass accretion rate as a function of radius & 0.5 \\
T_e & electron temperature & \ldots \\
t & power-law index for the electron temperature as a function of radius & \ldots \\
f & fraction of viscously dissipated energy that gets advected & 1 \\
\delta & fraction of viscous heating that goes directly to the electrons & 0.3 \\
r_{\text{min}} & minimum dimensionless radius of the advection region & 3 \\
r_{\text{max}} & maximum dimensionless radius of the advection region & $10^3$ \\
\theta_e & dimensionless electron temperature; $\frac{k T_e}{m_e c^2}$ & \ldots \\
\midrule
x_M & dimensionless synchrotron frequency; \autoref{eqn:DimensionlessSynchFreq} & \ldots \\
\nu_b & gyro frequency; $\frac{e B}{2 \pi m_e c}$ & \ldots \\
\nu_c & critical frequency below which synchrotron emission is optically thick & \ldots \\
\nu_p & ``peak'' critical synchrotron frequency at innermost radius; $\nu_p = \nu_c(r_{\text{min}})$ & \ldots \\
\nu_m & critical synchrotron frequency at outermost radius; $\nu_m = \nu_c(r_{\text{max}})$ & \ldots \\
\midrule
\gamma & ratio of specific heats; $\frac{8 + 5 \beta}{6 + 3 \beta}$ & 1.4 \\
\epsilon' & $\frac{1}{f} \left( \frac{5/3 - \gamma}{\gamma - 1} \right) = f^{-1} (1 + \beta)^{-1}$ & 0.5 \\
c_1 & $\frac{5 + 2 \epsilon'}{3 \alpha^2} \left( \sqrt{1 + \frac{18 \alpha^2}{\left( 5 + 2 \epsilon' \right)^2}} - 1 \right)$ & 0.5 \\
c_3 & $\frac{2 (5 + 2 \epsilon')}{9 \alpha^2} \left( \sqrt{1 + \frac{18 \alpha^2}{\left( 5 + 2 \epsilon' \right)^2}} - 1 \right)$ & 0.3 \\
s_1 & $1.42 \times 10^9 \sqrt{\frac{c_3}{\alpha c_1 (1 + \beta)}}$ & $1.5 \times 10^9$ \\
s_2 & $1.19 \times 10^{-13} x_M$ & \ldots \\
s_3 & \ldots & $1.05 \times 10^{-24}$ \\
b_1 & $3.16 \times 10^{19} \alpha^{-1} c_1^{-1}$ & $10^{20}$
\enddata
\tablecomments{A list of the parameters used for the SED model.  Certain parameters in the model take on the default values listed here, while others must either be specified as inputs (e.g., $m$, $\dot{m}$) or else are internally computed as part of the model (e.g., $T_e$, $t$).}
\end{deluxetable}

\clearpage

\section{Mass dependence of the Eddington ratio distribution function} \label{app:ERDFMassDependence}

As described in \autoref{sec:ERDF}, in this paper we take the ERDF to have a broken power-law functional form (see \autoref{eqn:ProbDistEdd}) with a power-law index $\alpha$ that evolves with both SMBH mass $M$ and redshift $z$.  Specifically, we adopt the redshift evolution prescription from \cite{Tucci_2017} (see \autoref{eqn:ERDFalpha}), and we add to it an evolution with SMBH mass (see \autoref{eqn:ERDFa}).  For the mass evolution of the ERDF power-law index, we choose a logistic function in $\log(M)$ such that low-mass SMBHs (i.e., those with masses below some value $M_0$) see a power-law index $a_{\text{lo}}$ while high-mass SMBHs (i.e., those with masses above $M_0$) see a power-law index $a_{\text{hi}}$.  The specific functional form of \autoref{eqn:ERDFa} ensures that $a(M)$ transitions smoothly between the low- and high-mass regimes, with a logarithmic width that is set by the parameter $\Delta$.

To determine the values of the ERDF parameters $a_{\text{lo}}$, $a_{\text{hi}}$, $M_0$, and $\Delta$, we rely on the observational constraints provided by \cite{Aird_2018}.  \cite{Aird_2018} determined the distribution of specific SMBH accretion rates $\lambda_s$ -- i.e., the accretion rate relative to the stellar mass of the galaxy, rather than to the mass of the SMBH -- by fitting a Bayesian mixture model to X-ray observations of ${\sim}10^5$ near-infrared-selected galaxies.  This sample includes a mix of star-forming, quiescent, and AGN-dominated galaxies, and it spans a range ${\sim}10^{8.5}$--$10^{11.5}$\,\msun in stellar mass and ${\sim}0.3$--4 in redshift.  We use two different prescriptions to convert from stellar mass to SMBH mass, corresponding to the two BHMF prescriptions described in \autoref{sec:BHMF}.  For our fiducial choice of the lower BHMF from \cite{Shankar_2009}, we adopt the stellar-to-SMBH conversion used by \cite{Aird_2018} themselves, which is given simply by $M_* = 500 M$; we use this fiducial prescription for all figures and values in this paper unless otherwise specified.  For the instances in which we quote a range of values corresponding to the lower and upper BHMFs, for the upper BHMF we convert from $\lambda_s$ to $\lambda$ using the same stellar-to-SMBH conversion as in \autoref{sec:BHMF} (i.e., \autoref{eqn:KHConversion}).

\begin{figure}
    \centering
    \includegraphics[width=1.00\textwidth]{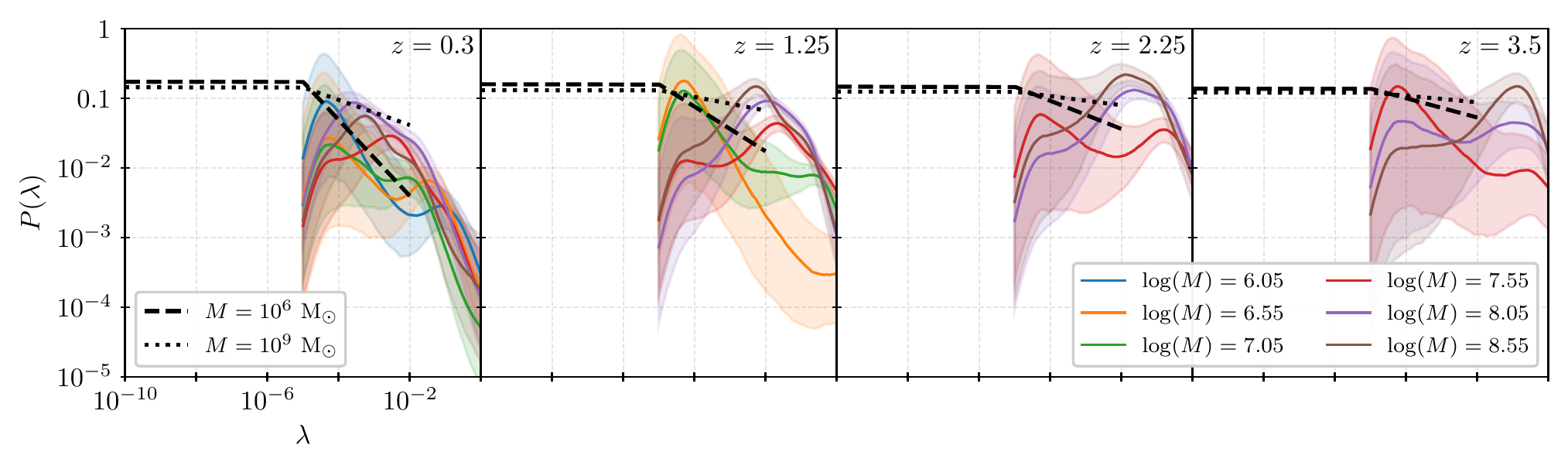}
    \caption{Comparison of the ERDF prescription used in this paper with the empirical modeling from \cite{Aird_2018}, with each panel showing a different choice of redshift.  In each panel, the colored lines and corresponding shaded regions show the constraints from \cite{Aird_2018} for a range of SMBH masses, as labeled in the legend at the lower right; the black dashed and dotted lines show our ERDF prescription (see \autoref{eqn:ProbDistEdd}) for example SMBH masses of $10^6$ and $10^9$\,\msun, respectively, as labeled in the legend at the lower left.  All panels share the same horizontal and vertical axis ranges, which are explicitly labeled in the left panel.}
    \label{fig:ERDF_fitting}
\end{figure}

Given the observed $P(\lambda)$ as a function of $M$ and $z$, we determine the best-fit ERDF parameters by minimizing the squared logarithmic differences between the \cite{Aird_2018} empirical model and \autoref{eqn:ProbDistEdd}.  We restrict our fitting to the region of parameter space between $-4 \leq \log(\lambda) \leq -1$, with the low-$\lambda$ cutoff determined by the observational limitations and the high-$\lambda$ cutoff determined by our interest in LLAGNs.  The resulting best-fit values for the fiducial case are $a_{\text{lo}} = 0.55$, $a_{\text{hi}} = 0.20$, $\log(M_0) = 7.5$, and $\Delta = 0.3$, and \autoref{fig:ERDF_fitting} shows a comparison between the best-fit ERDF and the \cite{Aird_2018} model.  For the upper BHMF prescription, the only parameter that changes is $M_0$, for which we find a best-fit value of $\log(M_0) = 7.8$.

\clearpage

\section{Analytic approximations based on volumetric scaling relations} \label{app:AnalyticApprox}

The behavior of $N(\theta_r,\sigma_{\nu})$ seen in \autoref{fig:source_counts_230GHz_taucut} and \autoref{fig:1D_slices} takes on an apparently simple structure, whose gross properties can be understood in terms of simple volumetric scaling relations.

\subsection{Analytic approximation for the population source counts} \label{sec:AnalyticPop}

For a static universe in which SMBHs are distributed uniformly, the number of SMBHs that could be spatially resolved at an angular resolution $\theta_r$ by a telescope with arbitrary sensitivity will be proportional to $\theta_r^{-3}$.  Similarly, the number of SMBHs that could be detected at a sensitivity $\sigma_{\nu}$ by a telescope with arbitrary angular resolution will be proportional to $\sigma_{\nu}^{-3/2}$.  A simple function that captures both limiting behaviors is

\begin{equation}
N(\theta_r, \sigma_{\nu}) \approx \left[ \left( \frac{\theta_r}{40 \text{ \uas}} \right)^3 + \left( \frac{\sigma_{\nu}}{1 \text{ Jy}} \right)^{3/2} \right]^{-1} . \label{eqn:SimpleFullApApprox}
\end{equation}

\noindent Here, we've chosen the normalization to be such that we would expect to see $\gtrsim$1 black hole shadows with angular sizes smaller than $\theta_r = 40$\,\uas and with flux densities less than $\sigma_{\nu} = 1$\,Jy -- approximately matching the values appropriate for the SMBH in M87 \citep{Paper3,Paper4} -- and that at this angular resolution and flux density sensitivity we would expect to see $N \approx 1$ black hole shadow.

Following these expectations, we fit a simple functional form to the source counts of

\begin{equation}
\mathcal{N}(\theta_r, \sigma_{\nu}) = \left[ \left( \frac{\theta_r}{\theta_{r,0}} \right)^{\gamma} + \left( \frac{\sigma_{\nu}}{\sigma_{\nu,0}} \right)^{\kappa} \right]^{-1} . \label{eqn:SimpleFunction1}
\end{equation}

\noindent The model parameters $\theta_{r,0}$, $\gamma$, $\sigma_{\nu,0}$, and $\kappa$ are determined by minimizing the squared logarithmic differences between $\mathcal{N}(\theta_r, \sigma_{\nu})$ from \autoref{eqn:SimpleFunction1} and the complete numerical evaluation of $N(\theta_{r},\sigma_{\nu})$ from \autoref{eqn:SourceCountInitial} (see \autoref{sec:NumberOfShadows}), assuming an observing frequency of 230\,GHz.  Both functions are evaluated on a $200 \times 200$ grid of $(\theta_{r},\sigma_{\nu})$ points, logarithmically spaced between $[10^{-2},10^2]$\,\uas in $\theta_{r}$ and between $[10^{-9},10^1]$\,Jy in $\sigma_{\nu}$.

\begin{figure}
    \centering
    \includegraphics[width=1.00\textwidth]{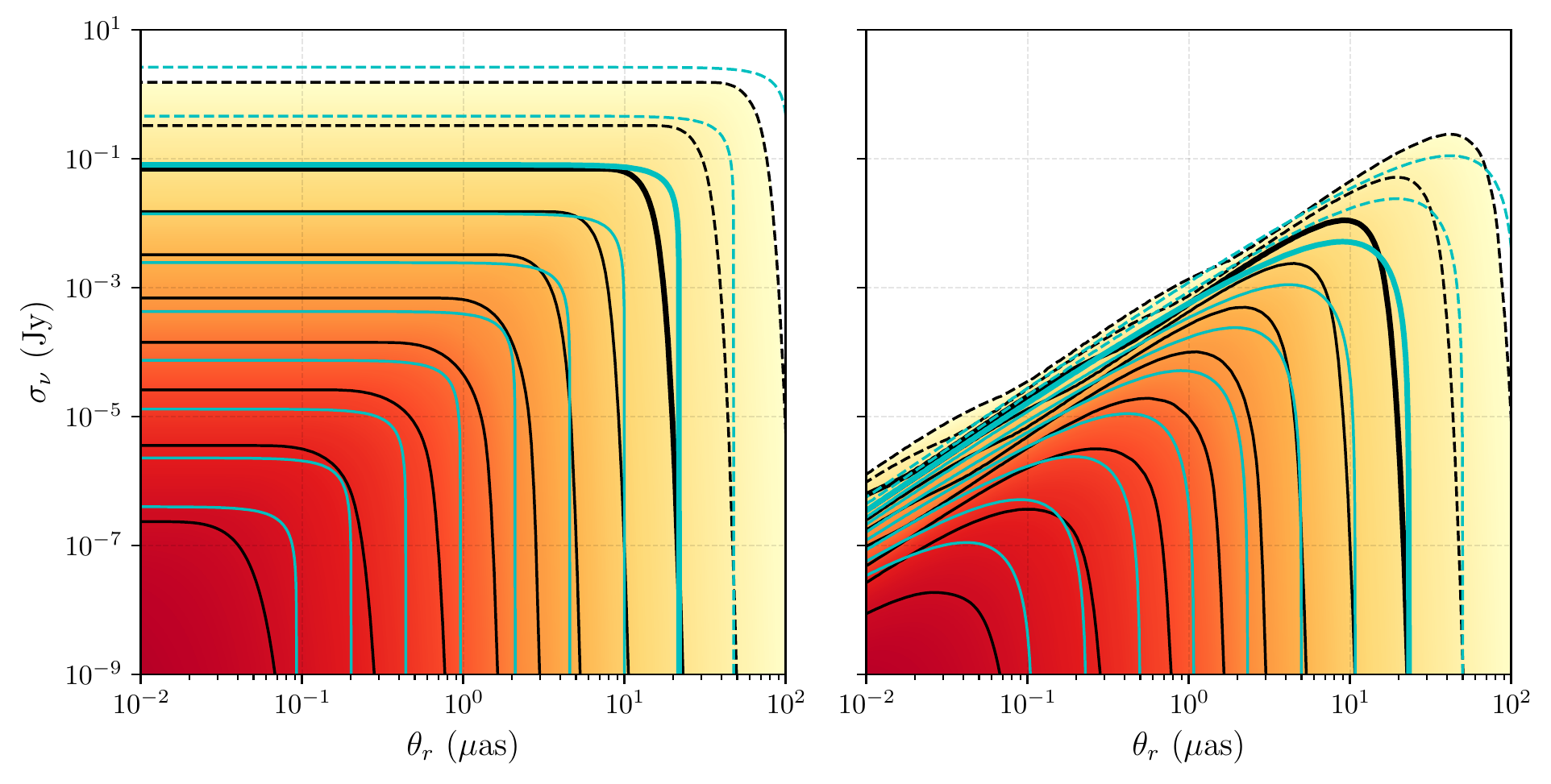}
    \caption{\textit{Left}: Same as the bottom panel of \autoref{fig:source_counts_230GHz_taucut}, but with the source counts predicted by the best-fitting analytic approximation (\autoref{eqn:SimpleFunction1}) overplotted in cyan contours. \textit{Right}: Same as \autoref{fig:source_counts_230GHz_taucut_rescut}, but with the source counts predicted by the best-fitting analytic approximation (\autoref{eqn:SimpleFunction2}) overplotted in cyan contours.}
    \label{fig:fitting_functions_simple}
\end{figure}

We find best-fit parameter values of $\theta_{r,0} = 21.8$\,\uas, $\gamma = 2.95$, $\sigma_{\nu,0} = 0.080$\,Jy, and $\kappa = 1.32$; this best fit is shown in the left panel of \autoref{fig:fitting_functions_simple}.  The power-law indices, $\gamma$ and $\kappa$, have best-fit values that are close the initial expectations (i.e., $\gamma = 3$ and $\kappa = 1.5$), indicating that the cosmological effects are not causing large deviations from simple volumetric scaling relations.  The normalization factors, $\theta_{r,0}$ and $\sigma_{\nu,0}$, are substantially different from the values in \autoref{eqn:SimpleFullApApprox}, in-line with the model's known underprediction of M87 (see \autoref{sec:M87}).  Overall, the best-fit \autoref{eqn:SimpleFunction1} provides a description of the source counts that deviates from the numerical computation by less than an order of magnitude across most of the $(\theta_{r},\sigma_{\nu})$ space.  Only for $\theta_r \lesssim 0.1$\,\uas and $\sigma_{\nu} \lesssim 10^{-7}$\,Jy does the analytic approximation deviate from the numerical computation by more than an order of magnitude in $N$.

\subsection{Analytic approximation for the interferometric source counts} \label{sec:AnalyticInf}

When observing with an interferometric baseline, the correlated flux density depends on the brightness temperature $T_b$ of the source emission.  For a source that is marginally resolved and which subtends a solid angle $\Omega$, the brightness temperature can be expressed as

\begin{eqnarray}
T_b & \approx & \frac{c^2 \sigma_{\nu}}{2 \nu^2 k \Omega} \nonumber \\
& \approx & \left( 1.64 \times 10^{10} \text{ K} \right) \left( \frac{\sigma_{\nu}}{1 \text{ Jy}} \right) \left( \frac{\theta_r}{40 \text{ \uas}} \right)^{-2} \left( \frac{\nu}{230 \text{ GHz}} \right)^{-2} .
\end{eqnarray}

\noindent Here, $\sigma_{\nu}$ represents the total flux density of the source and $\theta_r$ is its angular size on the sky. For synchrotron sources, self-absorption and energy equipartition are expected to limit $T_b$ to some maximum value of approximately $10^{11}$\,K \citep[][though see also \citealt{Kovalev_2016}]{Kellermann_1969,Readhead_1994}.  More specifically, the emitted brightness temperature should never exceed the electron temperature, which for our SED model described in \autoref{app:SEDmodel} does not go above ${\sim}7 \times 10^{10}$\,K (see the right panel of \autoref{fig:nup_and_Te_ve_m_mdot}).  Following the considerations from \autoref{sec:AnalyticPop} while also accounting for this brightness temperature limit, we can modify \autoref{eqn:SimpleFullApApprox} using an exponential cutoff to smoothly suppress the the high brightness temperature emission,

\begin{equation}
N(\theta_r, \sigma_{\nu}) \approx e^{-T_b/\left( 10^{10} \text{ K} \right)} \left[ \left( \frac{\theta_r}{40 \text{ \uas}} \right)^3 + \left( \frac{\sigma_{\nu}}{0.1 \text{ Jy}} \right)^{3/2} \right]^{-1} . \label{eqn:SimpleBaselineApprox}
\end{equation}

\noindent Here, $\theta_r$ should now be understood to represent a single-baseline angular resolution, and we have adjusted the $\sigma_{\nu}$ normalization to match the 230\,GHz flux density observed from M87 on long baselines \citep{Paper3}. We have set the brightness temperature cutoff to $10^{10}$\,K because we are selecting for SMBHs that are optically thin and which therefore should not typically saturate the brightness limit.  We note that magnetohydrodynamic simulations of the M87 system also exhibit brightness temperatures that peak between $10^{10}$\,K and $10^{11}$\,K \citep{Paper5}.

Following these expectations, we expand on the results of \autoref{sec:AnalyticPop} and fit a simple functional form to the source counts of

\begin{equation}
\mathcal{N}(\theta_r, \sigma_{\nu}) = \sum_{n=0}^{\infty} e^{-\left( T_b/T_{b,n} \right)^{\mu}} \left[ \left( \frac{\theta_r}{\theta_{r,n}} \right)^{3} + \left( \frac{\sigma_{\nu}}{\sigma_{\nu,n}} \right)^{3/2} \right]^{-1} . \label{eqn:SimpleFunction2}
\end{equation}

\noindent Here, we have fixed the exponents of the $\theta_r$ and $\sigma_{\nu}$ terms to the values expected from initial considerations and further motivated by the fitting results of \autoref{sec:AnalyticPop}.  We also incorporate the understanding from \autoref{sec:PhotonRingDecomp} into the scale-setting parameters $\theta_{r,n}$ and $\sigma_{\nu,n}$ for the $n$th sub-ring, which are defined to be

\begin{equation}
\theta_{r,n} = \theta_{r,0} e^{- n \pi} \quad \quad \quad \quad \sigma_{\nu,n} = \sigma_{\nu,0} e^{- 3 n \pi / 2} .
\end{equation}

\noindent We do not have an a priori expectation for the scaling behavior of the brightness temperature $T_{b,n}$ in each sub-ring, so we simply include it as an additional parameter,

\begin{equation}
T_{b,n} = T_{b,0} C^n .
\end{equation}

The model has five free parameters -- $\theta_{r,0}$, $\sigma_{\nu,0}$, $T_{b,0}$, $\mu$, and $C$ -- which we fit in the same manner described in \autoref{sec:AnalyticPop}.  We find best-fit parameter values of $\theta_{r,0} = 23.2$\,\uas, $\sigma_{\nu,0} = 0.17$\,Jy, $T_{b,0} = 2.2 \times 10^8$\,K, $\mu = 0.50$, and $C = 2.39$; this best fit is shown in the right panel of \autoref{fig:fitting_functions_simple}.  The fit quality is similar to that in \autoref{sec:AnalyticPop}, with the analytic source counts throughout most of the $(\theta_r,\sigma_{\nu})$ space agreeing to better than an order of magnitude with the numerical results.  The deviations become worse than an order of magnitude at small values of $\theta_r \lesssim 0.1$\,\uas and $\sigma_{\nu} \lesssim 10^{-7}$\,Jy, as well as wherever the source counts contain substantial contributions from $n > 0$ photon rings.

\clearpage

\section{Sampling function for a rotating baseline} \label{app:RotatingBaseline}

\begin{figure}
    \centering
    \includegraphics[width=1.00\textwidth]{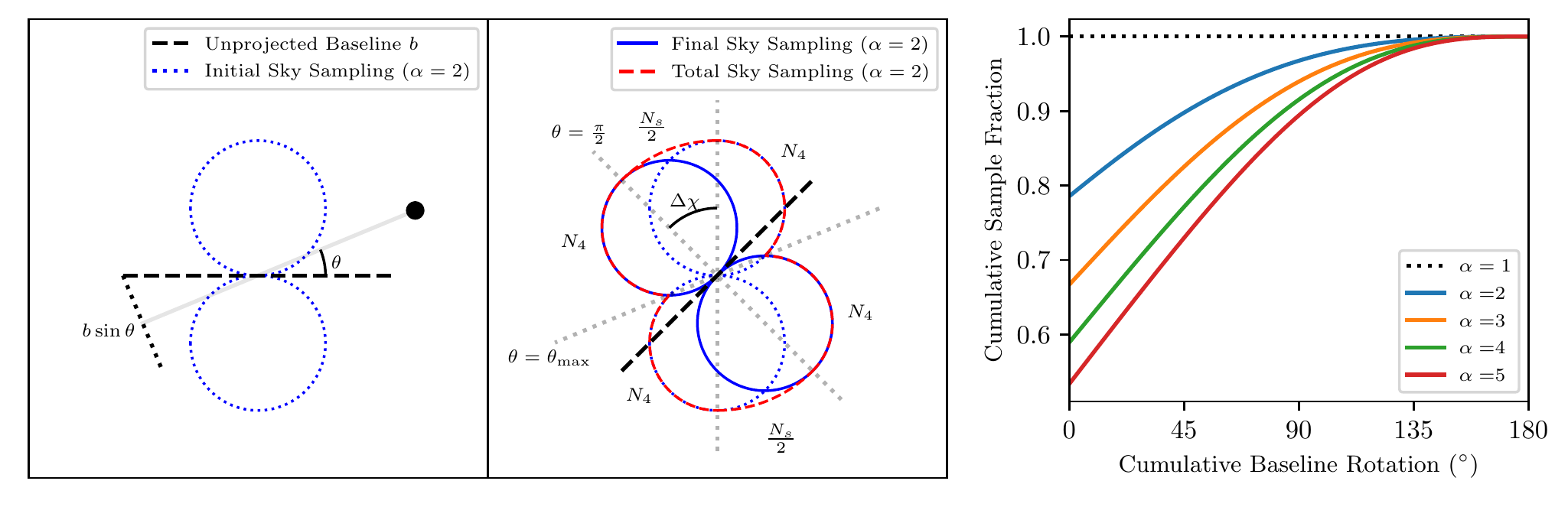}
    \caption{\textit{Left}: Schematic diagram of black holes sampled by a single baseline at one instant.  The dashed black line represents the baseline, the solid gray line represents the line of sight to an example black hole, the dotted black line represents the projection of the baseline perpendicular to that line of sight, and the dotted blue curve represents the sampling function (with radial distance from the baseline center proportional to the number of black holes sampled). \textit{Middle}: Same as left, but after the baseline evolves, rotating by an angle $\Delta \chi$, sweeping through the volume enclosed by the dashed red line. Dotted grey lines show the decomposition performed in \autoref{eqn:rotation_decomp}. Both left and middle diagrams correspond to a two-dimensional cross-section of a three-dimensional surface; for a black hole distribution that follows $\frac{d^3 N}{d\Omega d\theta_r} \propto \theta_r^{-\alpha}$ with $\alpha=2$, the instantaneous sampling surface is exactly a torus. \textit{Right}: Cumulative fraction of black holes sampled by sky density functions with increasingly steep dependence on resolution.}
    \label{fig:swept_sampling}
\end{figure}

Following \autoref{sec:BaselineProjection}, we would like to determine how the total number of sources that a particular baseline can resolve changes as that baseline rotates.  \autoref{eqn:BaselineRes} provides the angular resolution $\theta_r$ that a baseline of length $b$ (in units of the observing wavelength) has when viewed from a source at a sky position with polar angle $\theta$ as measured from the tip of the baseline (see the left panel of \autoref{fig:swept_sampling}); there is no $\phi$ dependence because sources are assumed to be distributed isotropically.  For any particular baseline, the number density of resolvable sources per unit solid angle can be written as

\begin{equation}
\frac{d^2 N}{d\Omega}(\theta,b) = \frac{1}{4\pi} \int_{\theta_r(\theta,b)}^{\infty} \frac{d^3 N}{d\Omega d \theta_r} \, \deriv \theta_r , \label{eqn:OnSkyNumberDensity}
\end{equation}

\noindent which can then be used as in \autoref{eqn:InstantaneousBaselineSampling} to determine the total number of resolvable sources for any fixed baseline $b$.  For the example calculations presented in this appendix, we will assume that the integrand of \autoref{eqn:OnSkyNumberDensity} follows a power law with index $\alpha$ and coefficient $A$,

\begin{eqnarray}
\frac{d^2 N}{d\Omega}(\theta,b) & = & \frac{1}{4 \pi} \int_{1/(b \sin\theta)}^\infty A \theta_r^{-\alpha} \, \deriv \theta_r \nonumber \\
& = & \frac{A}{4 \pi}\frac{(b \sin{\theta})^{\alpha-1}}{\alpha-1} .
\end{eqnarray}

\noindent The number of black holes instantaneously sampled is then given by an integral over solid angle, expressed in \autoref{eqn:InstantaneousBaselineSampling} but now evaluated explicitly for the power law:

\begin{eqnarray}
N_{\rm inst} &=& \int_0^{2\pi} \deriv \phi \int_0^\pi \sin\theta \frac{d^2 N}{d\Omega}(\theta,b) \, \deriv \theta \nonumber\\
&=& \frac{\sqrt{\pi}A b \Gamma(\frac{\alpha-1}{2})}{4 \Gamma(1+\frac{\alpha}{2})},
\label{eqn:AnalyticInstantaneousSampling}
\end{eqnarray}

\noindent where $\Gamma$ is the gamma function. In order to compare the cumulative sampling of different power laws, we must normalize so that after the baseline rotates by $180^\circ$, the number of black holes sampled is equal; this rotation corresponds to sweeping the largest projected spacing across the entire sky. Because the largest projected spacing is the only relevant quantity, we can normalize simply by requiring

\begin{equation}
N_{\rm total} = 4 \pi \frac{d^2 N}{d\Omega}(\theta=\frac{\pi}{2},b)
\end{equation}

\noindent for all $A$, $b$, and $\alpha$. We express the number of black holes sampled as a fraction of the total, removing the dependence on $A$ and $b$:

\begin{equation}
\frac{1}{N_{\rm total}}\frac{d^2 N}{d\Omega}(\theta) = \frac{(\sin\theta)^{\alpha-1}}{4 \pi} . \label{eqn:FractionalBHDensity}
\end{equation}

\noindent It is then straightforward to observe that the fraction of black holes in the sky instantaneously observed by a single baseline decreases with $\alpha$. However, it is less straightforward to compute the cumulative number of unique sources a baseline resolves over the course of some rotation through space, because the sources sampled at each baseline orientation are partially redundant with the source sampled at prior baseline orientations.  Put another way, the instantaneous sampling of black holes provided by a single baseline is given by a sweeping of the projected baseline through $\phi$, whereas the sampling over a change in orientation is a sweep through a new angle, which we call $\chi$. The angle $\chi$ is related to $\theta$ by

\begin{equation}
\chi = \tan^{-1}\left(\sin\phi \tan\theta\right) .
\end{equation}

\noindent $\chi$ is defined so that it aligns with $\theta$ when $\phi=\pi/2$. An illustration of $\chi$ and the redundant sampling are shown (for the $\phi=\pi/2$ cross-section) in the middle panel of \autoref{fig:swept_sampling}.

The rotation of the baseline in space can then be described by a change in the axial angle $\Delta\chi$. Computing \autoref{eqn:InstantaneousBaselineSampling} for the total sources sampled after this rotation (shown by the dashed red surface in the middle panel of \autoref{fig:swept_sampling}) can be simplified by breaking the integral into two regions: first, the longest projected baseline sweeps out a partial spheroid, while the rest of the baselines form arcs that intersect at the cusps of the dashed red line in \autoref{fig:swept_sampling}. We refer to the sources sampled by the partial spheroid as $N_s$ and those sampled by the cusped curves as $N_4$. As $\Delta \chi$ increases, $N_s$ increases and $N_4$ decreases. The cusped curves have a four-fold symmetry, so we integrate over a convenient curve (that between $\theta=\pi/2$ and $\theta = \pi$), and \autoref{eqn:InstantaneousBaselineSampling} reduces to

\begin{equation}
N(\Delta \chi) = 4 N_4(\Delta \chi) + N_s (\Delta \chi),
\label{eqn:rotation_decomp}
\end{equation}

\noindent where

\begin{equation}
N_4(\Delta\chi) = \int_0^\pi \deriv \phi \int_{\pi/2}^{\theta_{\text{max}}(\Delta\chi)} \sin\theta \frac{d^2 N}{d\Omega}(\theta) \, \deriv \theta
\end{equation}

\noindent is the contribution from each of the cusped curves,

\begin{equation}
N_s(\Delta\chi) = \frac{\Delta \chi}{\pi} \times 4 \pi \left[ \frac{d^2 N}{d\Omega}\left(\theta=\frac{\pi}{2}\right) \right]
\end{equation}

\noindent is the contribution from the partial spheroid, and

\begin{equation}
\theta_{\text{max}}(\Delta \chi) = \tan^{-1} \left( \frac{\tan\left( \pi - \frac{\Delta\chi}{2} \right)}{\sin\phi} \right)
\end{equation}

\noindent is the value of $\theta$ at the first leading cusp. The geometry of this computation is shown between the dotted gray lines in the middle panel of \autoref{fig:swept_sampling}.

The right panel of \autoref{fig:swept_sampling} shows the cumulative source sampling for several values of the power-law index $\alpha$ from \autoref{eqn:FractionalBHDensity}.  We use $\alpha=2$ for the schematic diagrams in \autoref{fig:swept_sampling} because the sampled surface around an instantaneous baseline in this case reduces to a torus; $\alpha = 4$ corresponds to a sky density of black holes that scales volumetrically (i.e., proportional to $\theta_r^{-3}$), which is closer to the actual behavior. For plausible values of $\alpha$, we find that the difference between a fully swept sampling of the sky and the instantaneous baseline sampling is not more than a factor of 2.

\end{document}